\documentclass[aps,preprint,showpacs,preprintnumbers,floatfix]{revtex4}

\usepackage{graphicx}
\usepackage{dcolumn}
\usepackage{bm}

\begin{document}
\preprint{UK/04-13}

\title{Study of pentaquarks on the lattice with overlap fermions}

\author{N.~Mathur~$^{a}$, F.X. Lee~$^{b,c}$, 
A. Alexandru~$^{a}$, C. Bennhold~$^{b}$, Y. Chen~$^{d}$, S.J.~Dong~$^{a}$,
T.~Draper~$^{a}$, I.~Horv\'ath~$^{a}$, K.F.~Liu~$^{a}$, S. Tamhankar~$^{a}$, 
J.B.~Zhang~$^{e}$}
\affiliation{
$^a$~Department of Physics \& Astronomy, University of Kentucky, Lexington, 
KY 40506, USA \\
$^b$~Center for Nuclear Studies, Department of Physics,
The George Washington University,  Washington, DC 20052, USA \\
$^c$~Jefferson Lab, 12000 Jefferson Avenue, Newport News, VA 23606, USA \\
$^d$~Institute of High Energy Physics, Academia Sinica, Beijing 100039, 
P.R. China \\
$^e$~CSSM and  Department of Physics, University of Adelaide,  
SA 5005, Australia}
\begin{abstract}
We present a quenched lattice QCD calculation of spin-1/2 five-quark
states with $uudd\bar{s}$ quark content for both positive and negative
parities. We do not observe any bound pentaquark state in these
channels for either $I = 0$ or $I =1$.  The states we found are
consistent with KN scattering states which are checked to exhibit the
expected volume dependence of the spectral weight.  The results are
based on overlap-fermion propagators on two lattices, $12^3\times 28$
and $16^3\times 28$, with the same lattice spacing of 0.2 fm, and pion
mass as low as $\sim$ 180 MeV.
\end{abstract}
\pacs{12.38.Gc, 14.20.Gk, 11.15.Ha}
\maketitle
\section{Introduction}
Since the reported discovery~\cite{expt0} two years ago of an exotic
5-quark resonance, named $\Theta^+(uudd\bar{s})$, with a mass of about
1540 MeV and a narrow width of less than 20 MeV, there has been a
rapid growth of interest in the subject. Eleven more
experiments have reported the observation of the
state~\cite{expt_yes}. It also stimulated the search for other
pentaquarks~\cite{expt_other}.  It should be pointed out, however,
that there are also ten experiments reporting negative
results~\cite{expt_no}. One has to wait for high statistics
experiments to clarify the situation in order to establish the exotic
state beyond doubt.

The strangeness quantum number of $\Theta^+$ is $S=+1$, but its
isospin and spin-parity assignments are undetermined by the experiments.
Based just on the valence quark content, the isospin could be 0, 1, or
2. The spin-parity could be ${1\over 2}^\pm$, ${3\over 2}^\pm$, or
higher.  The isospin would have to be established by discovering the
other charge states, while the spin-parity assignment will have to
await detailed measurements of decay angular distributions.

The experiments were inspired by the Skyrme model
prediction~\cite{soliton}, and the experimental discoveries have in
turn spawned intense interest on the theoretical side, with studies
ranging from chiral soliton and large $N_c$ models~\cite{soliton1},
quark models~\cite{qm}, KN phase-shift analysis~\cite{kn}, QCD sum
rules~\cite{qcdsr}, and recent lattice
calculations~\cite{csikor,sasaki,chiu}.
\section{Interpolating fields}
Unlike ordinary mesons ($q\bar{q}$) and baryons ($qqq$), pentaquarks
do not have a unique color structure aside from being a color
singlet. For a spin-1/2 pentaquark state of the type $uudd\bar{s}$,
there are two decay modes with different isospin content, $K^0p$ and
$K^+n$.  The simplest local interpolating field can be written as a
color-singlet configuration of a product of color-neutral meson and
baryon interpolation fields,
\begin{equation}
\chi_{1}^{\mp} = \epsilon^{abc}\left(u^{Ta} C \gamma_5 d^b \right)
\left[u^c \left( \bar{s}^e \gamma_5 d^e \right) \mp \left\{u
\leftrightarrow d\right\}\right],
\label{op1}
\end{equation}
where sum over all the color indices $\{a,b,c,e\}$ is implied.  The
minus sign is for isospin I=0 and plus sign for I=1 respectively.
A slight variation with a different color contraction is given by
\begin{equation}   \label{chi2}
\chi_{2}^{\mp} =
\epsilon^{abc}\left( u^{Ta} C \gamma_5 d^b \right)
\left[u^e \left( \bar{s}^e \gamma_5 d^c \right)
\mp \left\{u \leftrightarrow d\right\}\right],
\label{op2}
\end{equation}
where the color indices $e$ and $c$ are positioned differently.  Both
interpolation fields in Eq.~(\ref{op1}) and Eq.~(\ref{op2}) have been
used in a lattice calculation to study the pentaquark
$uudd\bar{s}$~\cite{csikor}.  Another possible $I = 0$ interpolation field
inspired by the diquark-diquark-antiquark picture is
\begin{eqnarray}
\chi_{3}^{\Gamma} &=& \epsilon^{gce}\epsilon^{gfh}\epsilon^{abc}
\left(u^{Ta} C \gamma_5 d^b \right) 
\left(u^{Tf} C\Gamma          d^h \right)\Gamma  C^{-1} \bar{s}^{Te}\nonumber\\
 &=& \epsilon^{abc}
\left(u^{Ta} C \gamma_5 d^b \right) 
\left(u^{Tc} C\Gamma d^e \right)\Gamma  C^{-1} \bar{s}^{Te}\nonumber\\
&& - \epsilon^{abc}
\left(u^{Ta} C \gamma_5 d^b \right) 
\left(u^{Te} C\Gamma d^c \right) \Gamma C^{-1} \bar{s}^{Te},
\label{op3}
\end{eqnarray}
where $\Gamma = {\hbox{\{S,A\}}} 
\equiv\{1, \gamma_{\mu}\gamma_{5}\}$, and  we used the 
antisymmetric tensor relation
$\epsilon^{gce}\epsilon^{gfh} = \delta^{cf}\delta^{eh} -
\delta^{ch}\delta^{ef}$ to obtain the second part of the equation. 
The interpolation field in Eq.~(\ref{op3}) with $\Gamma = \hbox{S}$
has been studied in Ref.~\cite{sasaki} and Ref.~\cite{chiu}.

These interpolation fields couple to states with
both parities. For $\chi_{3}$, as for the standard nucleon interpolation field,
the positive-parity state propagates in the forward (backward) time
direction in the upper (lower) Dirac components of the correlation
function, while the negative-parity state propagates backward (forward)
in the upper (lower) components.  On the other hand, 
for  $\chi_{1}$ and $\chi_{2}$,
the positive-parity state propagates in the forward (backward) time
direction in the lower (upper) Dirac component of the correlation
function, while the negative-parity state propagates backward (forward)
in the lower (upper) component.  In our calculations, we use both
upper and lower components to improve the statistics.

In general, since $\chi_{1}^{-}, \chi_{2}^{-}$ and  $\chi_{3}^{\Gamma}$
have the same quantum numbers ({\it e.g.} I = 0 and
$J^P = 1/2^{\pm}$) they should project out
the same states, albeit with different spectral weights. It is
interesting to note that they can be explicitly related. 
Indeed, despite their apparent different
color-spin structures, the KN type interpolation fields and that of
the diquark-diquark-antiquark type are related by a factor of
$\gamma_5$ and a Fierz re-arrangement which switches the roles of the
$u$ quark and $\bar{s}$ in Eq.~(\ref{op1}), leading to the following
expression 
\begin{eqnarray}
\hspace*{-0.6in}(u^{Ta} C\gamma_5 d^{b})\gamma_5 u^c (\bar{s}^e\gamma_5 d^e) 
=&&
\,\,\,\,{1\over 4}(u^{Ta} C\gamma_5 d^{b})(u^{Tc} Cd^{e})C^{-1}\bar{s}^{Te}
\nonumber \\ & &
+{1\over 4}(u^{Ta} C\gamma_5 d^{b})(u^{Tc} C \gamma_\mu d^{e})\gamma_\mu C^{-1}\bar{s}^{Te}
\nonumber \\ & &
-{1\over 8}(u^{Ta} C\gamma_5 d^{b})(u^{Tc} C \sigma_{\mu\nu} d^{e})\sigma_{\mu\nu} C^{-1}\bar{s}^{Te}
\nonumber \\ & &
+{1\over 4}(u^{Ta} C\gamma_5 d^{b})(u^{Tc} C \gamma_\mu\gamma_5 d^{e})\gamma_\mu\gamma_5 C^{-1}\bar{s}^{Te}
\nonumber \\ & &
+{1\over 4}(u^{Ta} C\gamma_5 d^{b})(u^{Tc} C \gamma_5 d^{e})\gamma_5 C^{-1}\bar{s}^{Te}.
\label{fiertz}
\end{eqnarray}
In this expression, a sum over dummy indices $\mu$ and $\nu$ is
implied, so the right-hand-side has 16 terms. We use the $\gamma$--matrix
relation $\{\gamma_{\mu}, \gamma_{\nu}\} = 2 \delta_{\mu\nu}$.
We see that the first term on the right hand side is just the first
term in the second part of Eq.~(\ref{op3}). A similar Fierz transform
of $\chi_2$ in Eq.~(\ref{chi2}) will lead to the second term in the
second part of Eq.~(\ref{op3}).  Since $\gamma_5$ multiplication
reverses parity, the extra $\gamma_5$ factor in front of the $u$ quark
on the left hand side serves to match the explicit parity of both
sides.  Therefore, schematically one can write,
\begin{eqnarray}
\lefteqn{
\gamma_5\times  (\mbox{KN interpolation field})} \nonumber \\
& = {1\over 2}\times (\mbox{diquark-diquark-antiquark interpolation field})\nonumber\\
& +\, \mbox{other terms},
\label{f1}
\end{eqnarray}
where the KN interpolation field is either $\chi_{1}$ or  $\chi_2$.
More precisely, 
\begin{equation}
  \gamma_5 \left( \chi_{1}^{-} - \chi_{2}^{-} \right)  =  
  {1\over 2} \left(\chi_{3}^{S} + \chi_{3}^{A}\right). 
\end{equation}
The fact that the KN interpolation field and the
diquark-diquark-antiquark interpolation field are directly related
further enhances the argument that both interpolation fields couple to the 
same physical spectrum with different strengths. This exercise also suggests
other possibilities for the diquark-diquark-antiquark interpolation
fields for pentaquarks. In view of this, it is not surprising that the
ground state results using KN type interpolation fields~\cite{csikor}
largely agree with those~\cite{sasaki} using the
diquark-diquark-antiquark type.  But it is a puzzle that
Ref.~\cite{chiu}, using the same diquark-diquark-antiquark type
interpolation field, produces qualitatively different results than those
of Ref.~\cite{csikor} and~\cite{sasaki}. Although Ref.~\cite{chiu} uses
the overlap fermion, which has better chiral properties than the Wilson 
fermion adopted in Refs.~\cite{csikor} and~\cite{sasaki}, one would not 
expect a qualitative difference in these calculations between the two 
fermion formalisms for the pion masses larger than 440 MeV. We will compare 
our results with these lattice calculations later.

The correlation function is obtained by Wick-contractions of all possible
quark pairs.  In the case of the interpolating field in
Eq.~(\ref{op1}), it has four terms. Due to isospin symmetry 
in the $u$ and $d$ quarks, the two diagonal terms are equal, and so are the two
cross terms. The zero-momentum correlation function without the
upper(lower) Dirac component projection reads
%
\begin{eqnarray}
G_{5q}(t) = \sum_{ \vec{x}}\langle\, \chi(x)\, \bar{\chi}(0)\,\rangle
= \sum_{ \vec{x}}\langle\, \chi(x)\, \bar{\chi}(0)\,\rangle_{diag} 
\pm \,\,\langle\, \chi(x)\, \bar{\chi}(0)\,\rangle_{cross},
\label{5q}
\end{eqnarray}
where the diagonal contribution is given by,
\begin{eqnarray} 
\sum_{ \vec{x}}\langle\, \chi(x)\, \bar{\chi}(0)\,\rangle^{\alpha\beta}_{diag} 
&=&  
2\,\sum_{ \vec{x}} \epsilon_{abc}\epsilon_{a^{\prime}b^{\prime}c^{\prime}} \, 
\times \nonumber \\ 
&&\hspace*{-1.4in} 
{\biggr\{} 
\, S^{aa^{\prime}}_{\alpha\beta}(u) {\hbox{Tr}} \left[S(d)S^{\dagger}(s) \right]
{\hbox{tr}} \left[{\underline{S}}^{bb^{\prime}}(d)S^{cc^{\prime}}(u)\right]\nonumber \\ 
&&\hspace*{-1.3in} \, - \, S^{aa^{\prime}}_{\alpha\beta}(u) {\hbox{tr}} 
\left[ {\cal{P}}^{bb^{\prime}}(d,s,d)\underline{S}^{cc^{\prime}}(u) \right] \nonumber \\ 
&&\hspace*{-1.3in} \, + \, \left[ S^{aa^{\prime}}(u)\underline{S}^{bb^{\prime}}(d)S^{cc^{\prime}}(u)\right]_{\alpha\beta}
{\hbox{Tr}} \left[S(d)S^{\dagger}(s) \right] \nonumber \\
&&\hspace*{-1.3in} \, - \, \left[ S^{aa^{\prime}}(u){\cal{\underline{P}}}^{bb^{\prime}}(d,s,d)S^{cc^{\prime}}(u)\right]_{\alpha\beta}
{\biggr\}}\,,
\end{eqnarray}
and the cross contribution is
\begin{eqnarray} 
\sum_{ \vec{x}}\langle\, \chi(x)\, \bar{\chi}(0)\,\rangle^{\alpha\beta}_{cross} 
&=&  
2\,\sum_{ \vec{x}} \epsilon_{abc}\epsilon_{a^{\prime}b^{\prime}c^{\prime}} \, \times \nonumber \\ 
&&\hspace*{-1.4in} 
{\biggr\{} 
 \, {\cal{P}}^{aa^{\prime}}_{\alpha\beta}(d,s,u){\hbox{tr}} \left[{\underline{S}}^{bb^{\prime}}(d)S^{cc^{\prime}}(u)\right]\nonumber \\ 
&&\hspace*{-1.3in}  +  \,\left[ S^{aa^{\prime}}(u)\underline{S}^{bb^{\prime}}(d){\cal{P}}^{cc^{\prime}}(d,s,u)\right]_{\alpha\beta}\nonumber \\
&&\hspace*{-1.3in} + \, \left[ {\cal{P}}^{aa^{\prime}}(d,s,u)\underline{S}^{bb^{\prime}}(d)S^{cc^{\prime}}(u)\right]_{\alpha\beta}\nonumber \\
&&\hspace*{-1.3in}  -  \,\left[ S^{aa^{\prime}}(d){\cal{\underline{P}}}^{bb^{\prime}}(d,s,u)S^{cc^{\prime}}(u)\right]_{\alpha\beta}
{\biggr\}}.
\end{eqnarray}
In the above expression, $S \equiv S_q(x,0)$ is the fully-interacting quark propagator,  
$\underline{S} \equiv  (\tilde{C} S \tilde{C}^{-1})^{T}$ 
with $\tilde{C} = C\gamma_{5}$ and ${\cal{P}}(u,s,d) =
S(u){S}^{\dagger}(s)S(d)$.  The trace on
spin-color is denoted by `Tr' while `tr' represents the trace on spin only.
In Eq.~(\ref{5q}) plus sign is for isospin I = 0 and minus sign for 
I = 1 respectively.

\section{Results and Discussion}
\begin{figure}
\includegraphics[height=6.0cm]{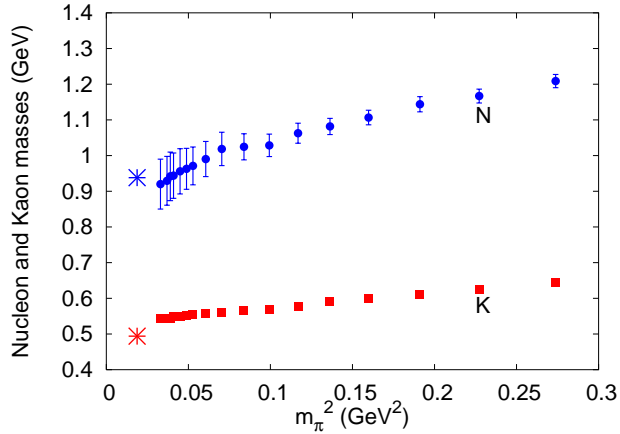}
\caption{Nucleon and kaon masses as a function of $m_{\pi}^{2}$ for 3.2 fm 
lattice. Experimental values are represented by the star symbols.}
\label{KN_mass}
\end{figure}

Our results below are obtained on two lattices, $12^3\times 28$, and
$16^3\times 28$, using the Iwasaki gauge action~\cite{Iwasaki85} and
the overlap fermion action~\cite{neu98}.  The lattice spacing of $a =
0.200(3)$ fm was determined from $f_\pi$~\cite{chlog04} for both
lattices, so the box size is L = 2.4 fm and L = 3.2 fm, respectively.
Our quenched quark propagators cover a wide range of quark masses,
corresponding to pion mass from 1293(20) MeV down to 182(8) MeV. The strange
quark mass is set by the $\phi$ meson, and corresponds to a pseudoscalar
 mass $m_\pi \sim 760$ MeV. In Fig. \ref{KN_mass}, we plot the nucleon 
and the kaon masses as a function of $m_{\pi}^{2}$. A naive linear 
extrapolation to the physical pion mass yields a kaon mass which is 
within $\sim 7\%$ of the corresponding experimental value (star symbol).
The Iwasaki gauge action is an ${\cal{O}}(a^2)$
renormalization-group improved action which allows the use of
relatively coarse lattices without suffering from large discretization
errors.  The overlap fermion action preserves exact chiral symmetry on
the lattice, and thus it has no ${\cal{O}}(a)$ error. One further
finds that ${\cal{O}}(a^2)$ errors are small for the meson
masses~\cite{dll00} and renormalization
constants~\cite{ddh02,kfliu02,kfliu03}.  Owing to the relatively
gentle critical slowing down~\cite{ddh02,kfliu02}, it allows us to work
at unprecedented small quark masses.  The relatively large box
size ensures that the finite-volume errors are under control.  At our
lowest pion mass, the finite-volume error is estimated to be $\sim$
2.7\%~\cite{chlog04}.  We have used the combination of Iwasaki gauge
action and overlap fermion action with local-local correlators
in a number of recent studies,
including chiral logs~\cite{chlog04} and baryon excited
states~\cite{roper04}. 
 To handle the excited states, we use the recently 
developed constrained-curve-fitting algorithm -- the {\it sequential 
empirical Bayes method}~\cite{fitting04}. The two ensembles studied here
contain 80 gauge configurations each.


\begin{figure*}[htb!]
\includegraphics[height=5.5cm]{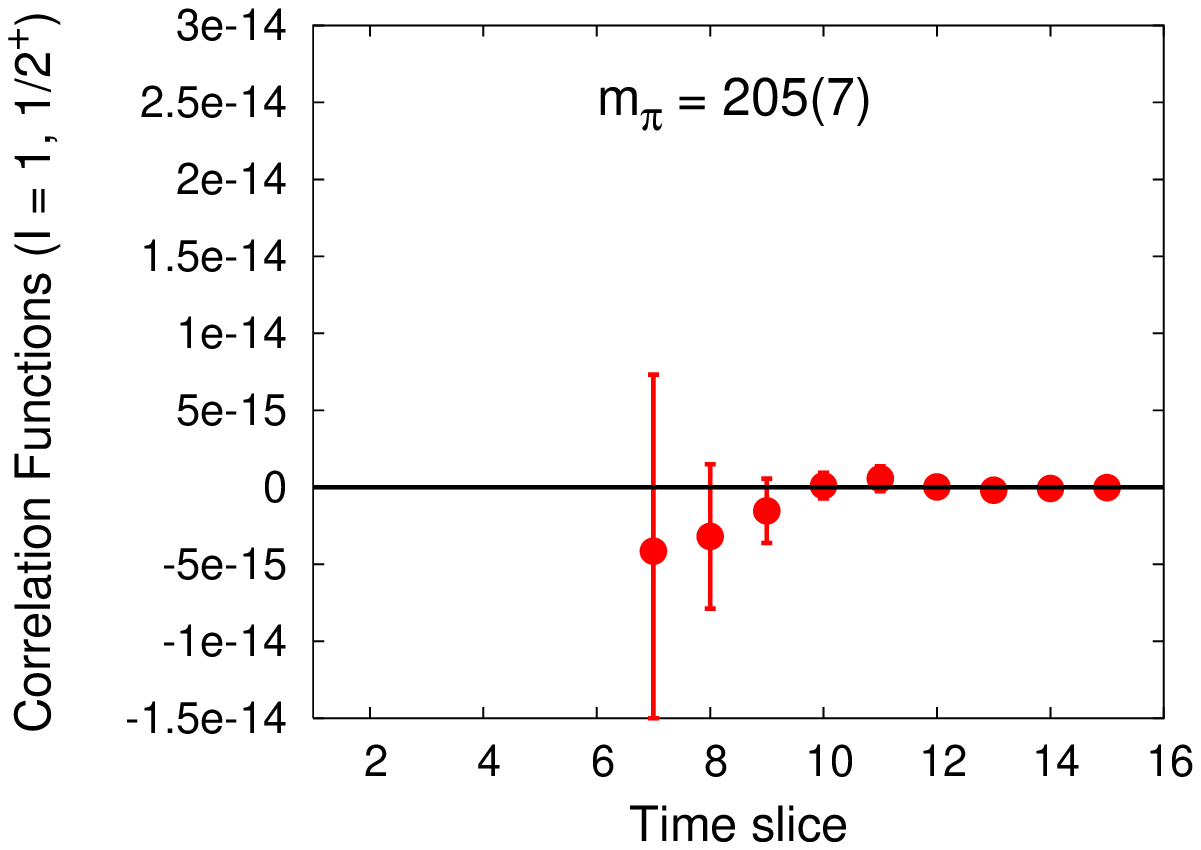}
\includegraphics[height=5.5cm]{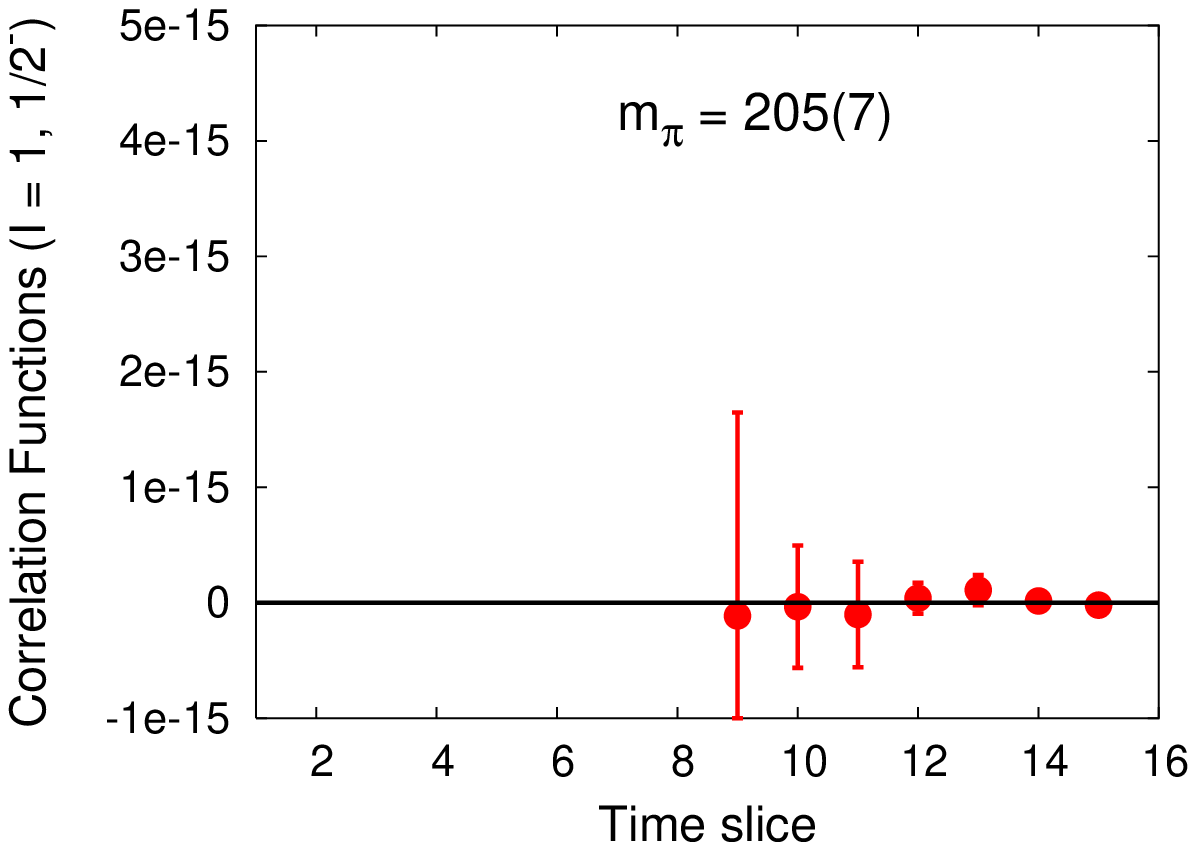}
\\
\includegraphics[height=5.5cm]{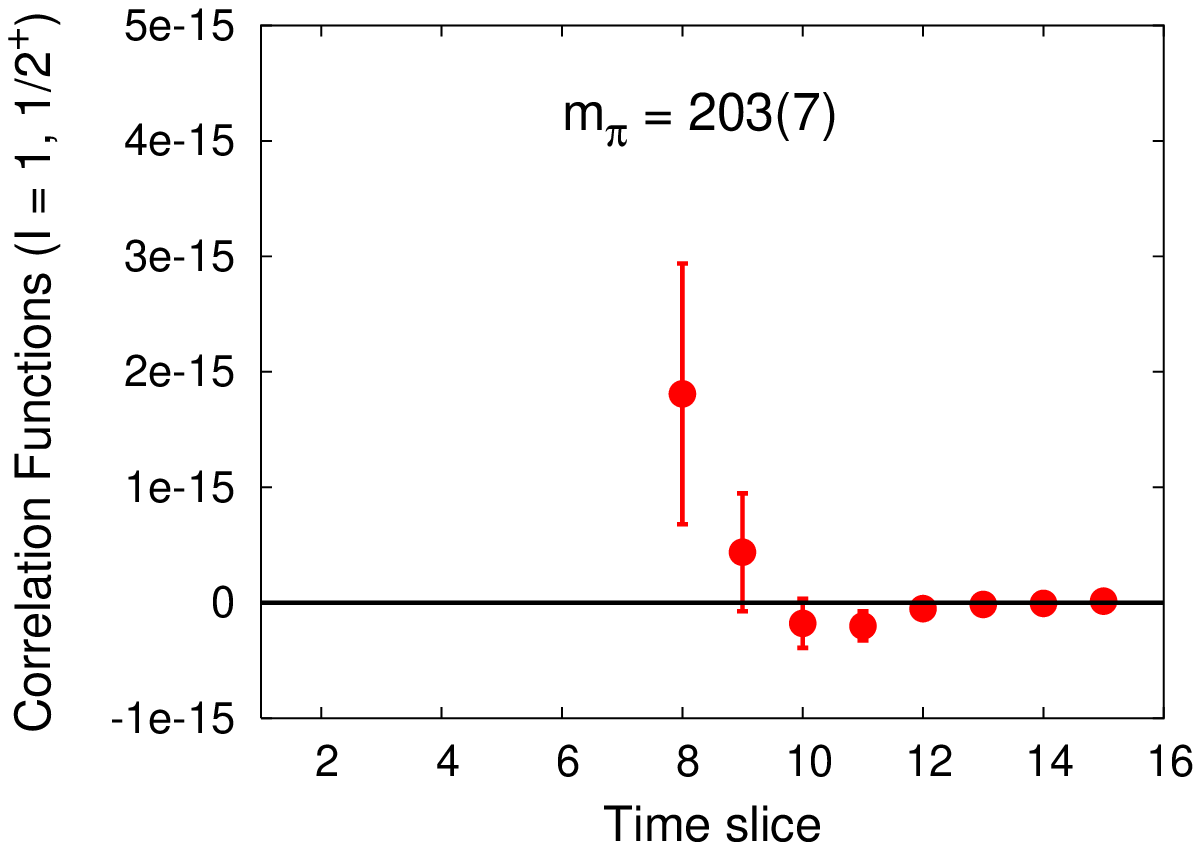}
\includegraphics[height=5.5cm]{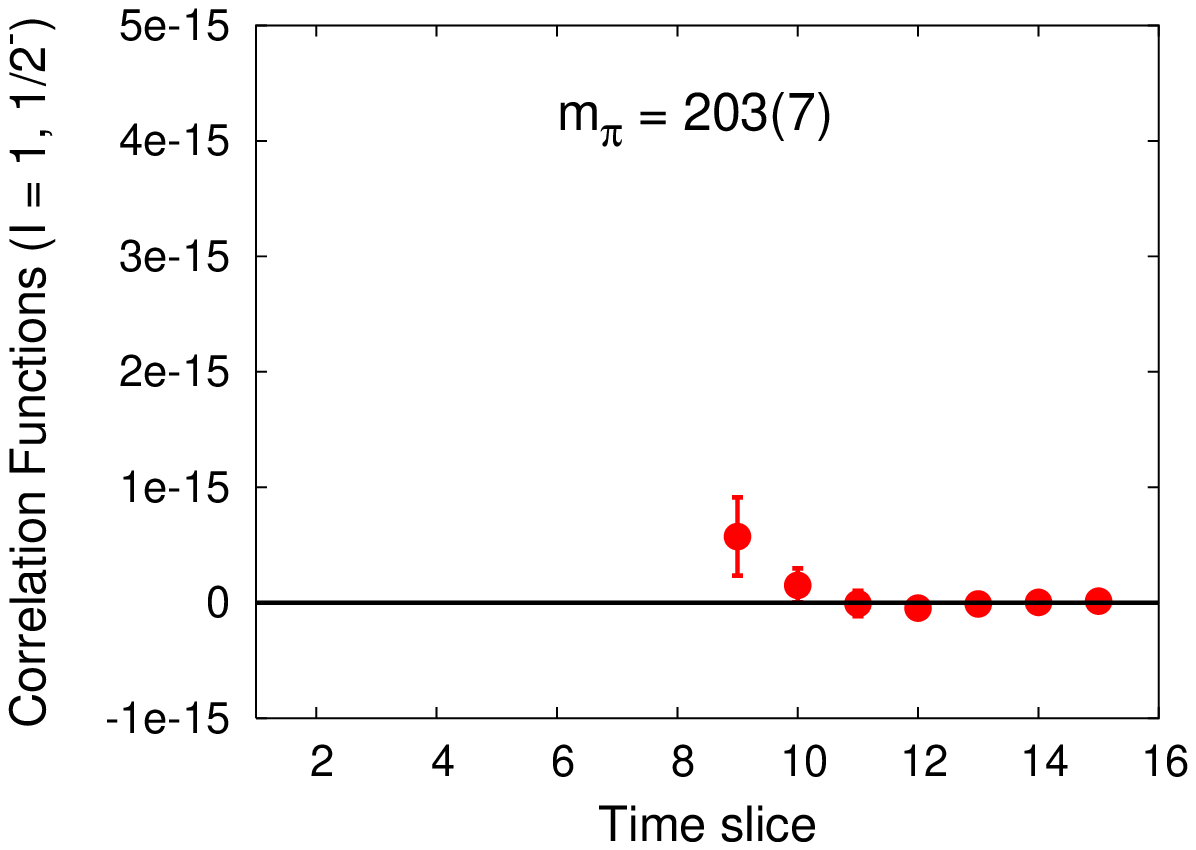}
\caption{Correlation functions for the positive parity (first column) 
and the negative parity (second column) for 2.4 fm (first row) and 3.2 fm
(second row), respectively (at a pion mass around 200 MeV).  One
should notice that the scale in the top left figure is different than
the others. This shows that for the positive parity channel the ghost
state is more prominent for the smaller lattice, while no ghost state
is detected for the negative parity channel.}
\label{cfun1}
\end{figure*}

\subsection{KN scattering states}
It is worth noting that the entire five-quark spectrum 
is contained in the correlation function in Eq.~(\ref{5q}), 
including a tower of KN scattering states
and possible bound pentaquarks.
Considering the presence of a pentaquark state, the above correlation
function can be written as a sum of exponentials
\begin{equation}
{G_{5q}(t)} = W_{pen} e^{-m_{pen}t} + \sum_i W_i e^{-E_{KN}^i t} + ...
\label{fit_5q}
\end{equation}
It should be stressed that the ordering of possible pentaquark states
and KN scattering states with energies $E_{KN}^i$ in
Eq.~(\ref{fit_5q}) is not known {\it a priori}, and must be determined
by fitting the data.  Furthermore, one needs to discern the nature of
the fitted states in order to distinguish if they are KN scattering
states or bound pentaquark states.  The parameter $E_{KN}^i$ denotes
the KN two-particle energies with zero total momentum. Since the KN
interaction is relatively weak, their values are expected to be near
the two-particle threshold energy defined as
\begin{equation}
E_{K}(p) + E_{N}(p)=\sqrt{m_K^2+p_{K}^2}+\sqrt{m_N^2+p_{N}^2}.
\label{E_KN}
\end{equation}
We use the discrete momentum available on the lattice:
 $p_{K}(n)=\sqrt{n}(2/a)\sin(\pi/L)$ for the kaon 
and $p_{N}(n) = \sqrt{n}(1/a)\sin(2\pi/L)$
 for the nucleon, where $n=0,1,2,\cdots$.
 As an example, the KN threshold
 energies corresponding to the first few discrete momenta 
are given in Table~\ref{tab_KN}. On our lattice, the energies calculated
from the above forms of 
lattice momenta deviate about 2-3\% from the corresponding energies
calculated from $p_{K} =p_{N} = 2\pi/L$.

We will refer to these discrete states as $p=0$, $p=1$ for $n = 0, n=1$,
and so on, keeping in mind that their actual values are given in
Table~\ref{tab_KN}.  One can see the expected down-shift of the states
on the larger volume.  The spacing between neighboring states
decreases as $n$ increases and the momentum states are somewhat more
packed for heavy quark cases.  The discrete KN scattering states play
different roles in the positive-parity and negative-parity channels.
In the positive-parity channel, they are in a relative $P$-wave, so
the spectrum starts at $p=1$, which has a raised threshold of 1.80 GeV
at $m_\pi=182$ MeV, while in the negative-parity channel, they are in
$S$-wave and the spectrum starts at zero momentum with a threshold of 1.47
GeV.  If there exists a pentaquark near 1.54 GeV, it would lie below
(above) the $KN$ threshold in the positive (negative) parity
channel. This means that it would be much easier to extract it in the
positive-parity channel than in the negative-parity channel.

\begin{table}
\caption{The KN threshold energies corresponding to the first 
few discrete momenta at three different pion masses are 
given for the two lattices. 
In each block, the first column is for the $12^3\times 28$ lattice, 
the second column  
$16^3\times 28$ lattice.}
\vspace*{0.2in}
\label{tab_KN}
\begin{tabular}{c|cc|cc|cc}
\multicolumn{1}{c|}{$n$\,}  & 
\multicolumn{2}{c|}{$E_{KN}$ (GeV)}  & 
\multicolumn{2}{c|}{$E_{KN}$ (GeV)}  & 
\multicolumn{2}{c }{$E_{KN}$ (GeV)}  \\ 
\multicolumn{1}{c|}{         }  & 
\multicolumn{2}{c|}{($m_\pi=182$ MeV)} & 
\multicolumn{2}{c|}{($m_\pi=438$ MeV)} & 
\multicolumn{2}{c}{($m_\pi=692$ MeV)} \\
 \hline
  0\,  &\, 1.47  & 1.47 &\, 1.77 & 1.77 &\, 2.16 & 2.16  \\
  1\,  &\, 1.80  & 1.67 &\, 2.06 & 1.94 &\, 2.41 & 2.30  \\
  2\,  &\, 2.07  & 1.85 &\, 2.30 & 2.10 &\, 2.63 & 2.44  \\
  3\,  &\, 2.31  & 2.00 &\, 2.52 & 2.24 &\, 2.83 & 2.56  \\
\end{tabular}
 \end{table}
The experimental values for the KN scattering length (volume) are
given in Table~\ref{tab_len}. The numbers are taken from
Ref.~\cite{phase}.  The interaction is very weak in the I=0 channel
(with a small attraction in the $P$-wave), and is slightly repulsive
in the I=1 channel.
\begin{table}
\caption{Experimental values for the KN scattering length (volume).}
\label{tab_len}
\begin{tabular}{c|cc}
scattering      &  I=0             &  I=1   \\
\hline
S-wave (fm)     &  $0.0\pm 0.03$   &  $-0.32\pm 0.02$   \\
P-wave (fm$^3$) &  $0.08\pm 0.01$  &  $-0.16\pm 0.1$   \\
\end{tabular}
 \end{table}
The $S$-wave scattering length on the lattice is related to the energy shift 
in a finite box of length $L$ by~\cite{luscher}
\begin{eqnarray}  \label{scatter-length}
E_{KN}^{\mbox{int}}-(E_K+E_N)=-{2\pi a_0 \over \mu_{KN}L^3}
\left[ 1+ c_1{a_0\over L} + c_2{a_0^2\over L^2}\right] + \,{\cal{O}}(L^{-6}),
\label{shift}
\end{eqnarray}
where $a_{0}$ is the $S$-wave scattering length and $\mu_{KN}$ 
is the reduced mass of the KN system.
Using the experimental numbers and neglecting $c_1$ and $c_2$, the
estimated energy shift for $I =1$ is about 8 MeV on our $16^3\times
28$ lattice,\, and about 18 MeV on the $12^3\times 28$ lattice.  As it will
become clear later, these values are consistent with our results.  One
could turn this argument around and use our simulation results to
extract information on the KN scattering~\cite{Chun_Liu}, a subject
outside the focus of this work.

\begin{figure*}[htb!]
\vspace*{-0.3in}
\includegraphics[width=7.5cm,height=3.8cm]{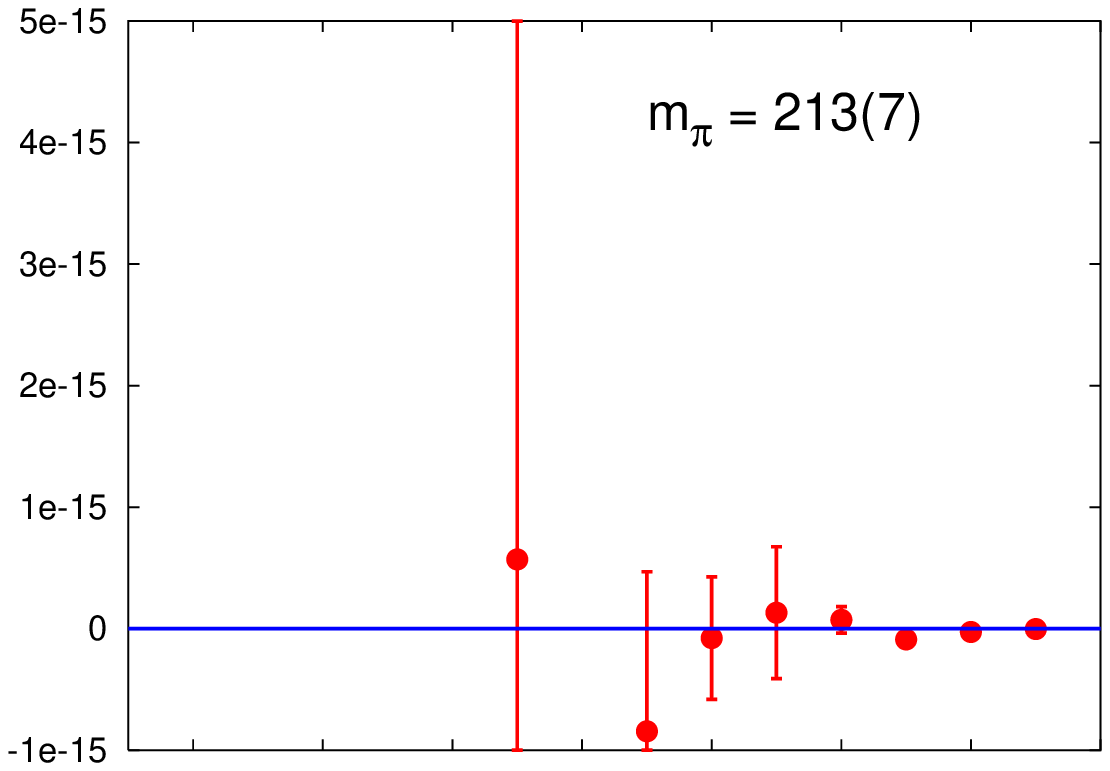}
\includegraphics[width=7.5cm,height=3.8cm]{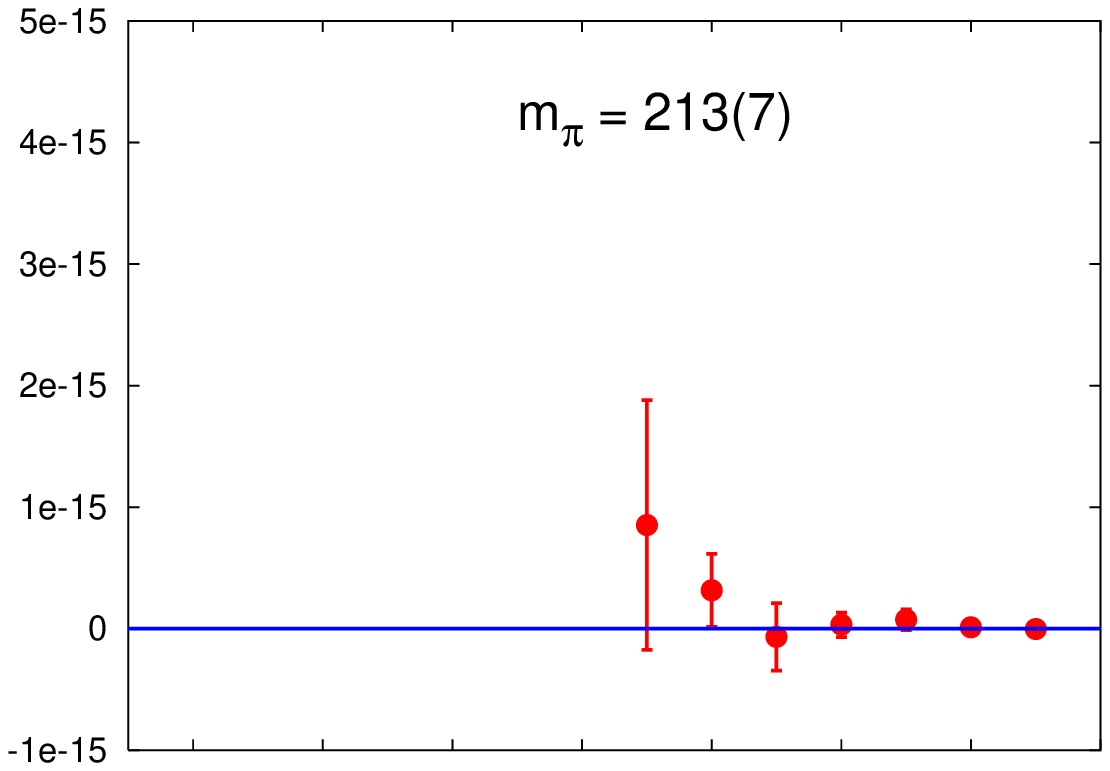}\\
\vspace*{-0.22in}
\includegraphics[width=7.5cm,height=3.8cm]{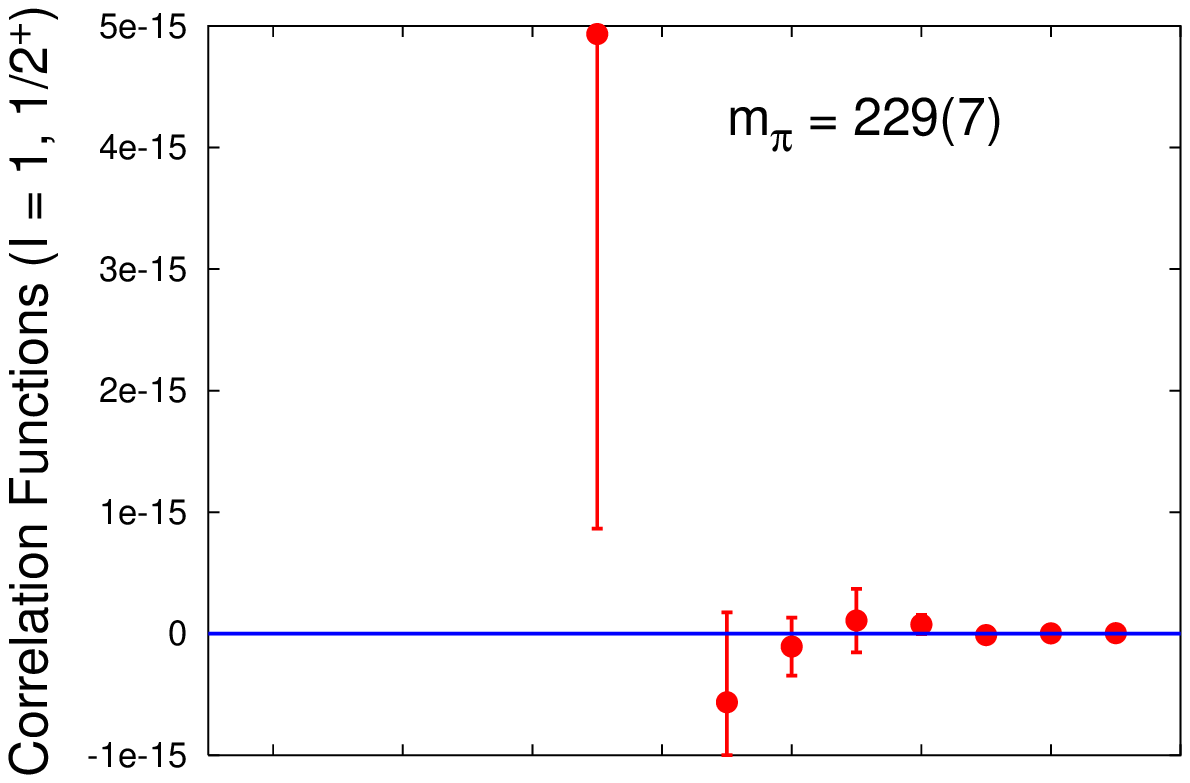}
\vspace*{-0.22in}
\includegraphics[width=7.5cm,height=3.8cm]{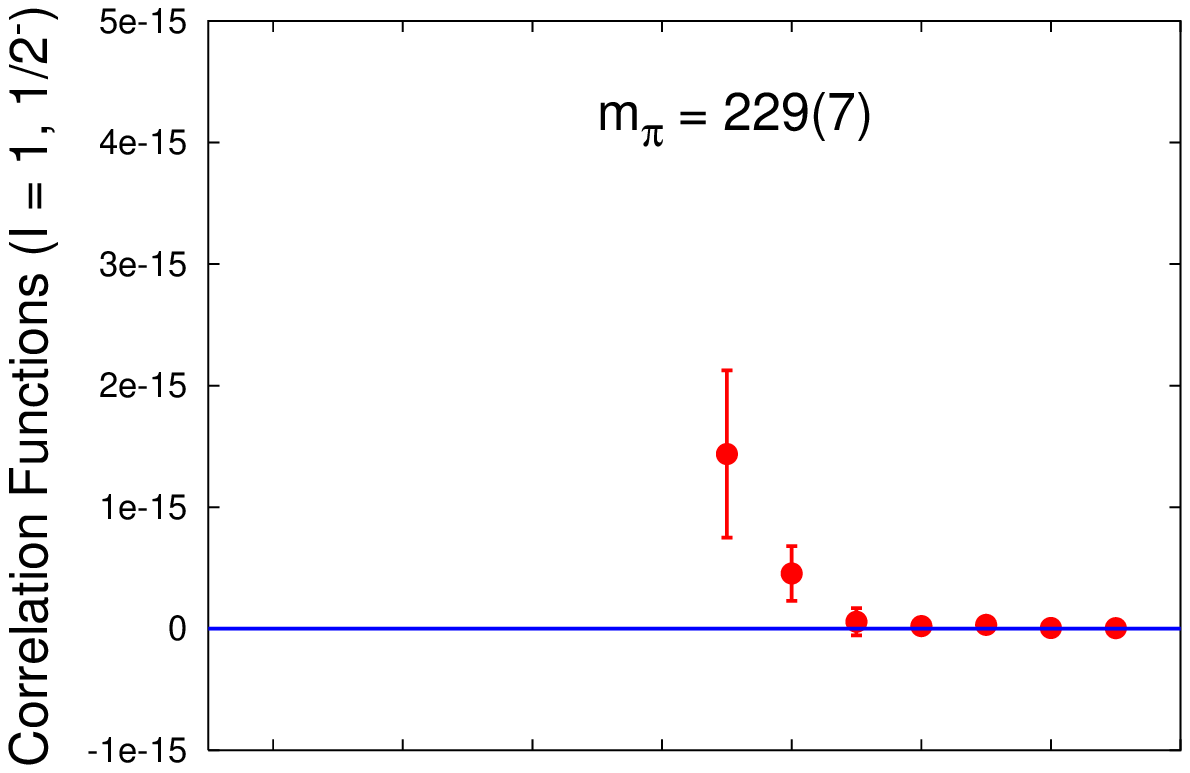}\\
\includegraphics[width=7.5cm,height=3.8cm]{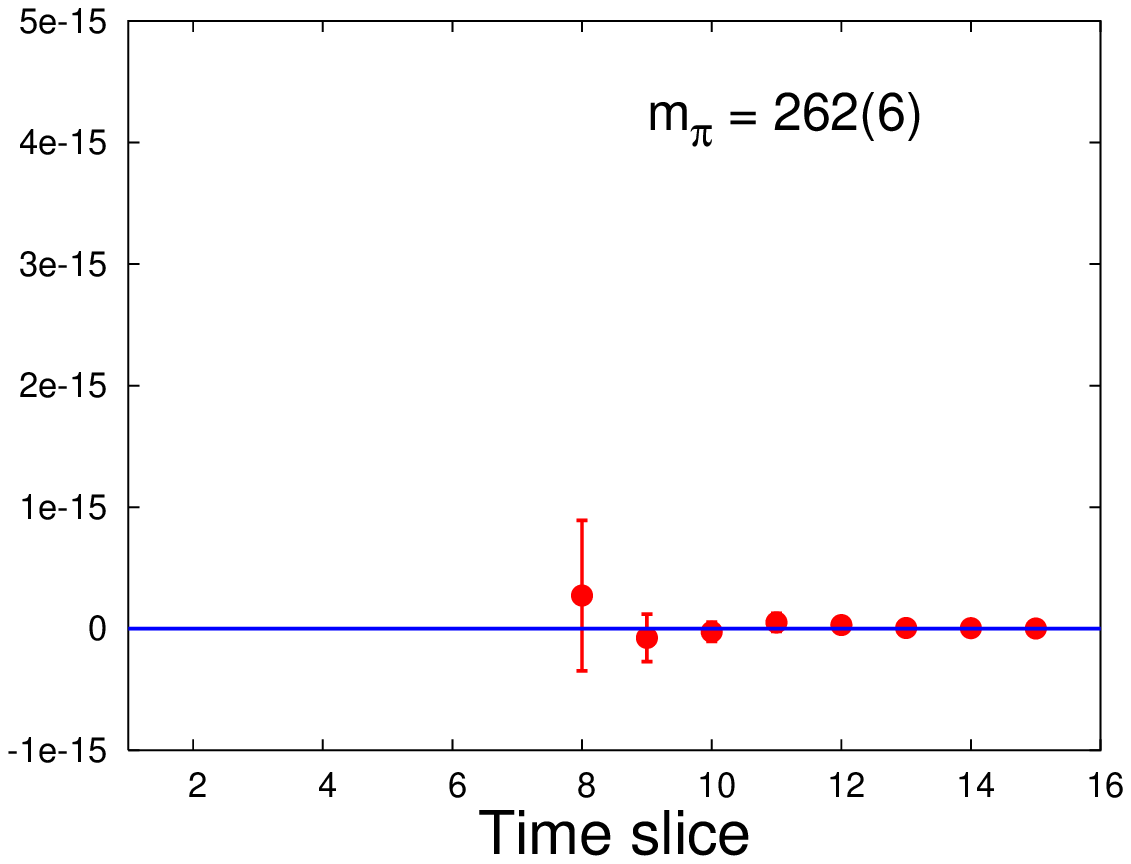}
\includegraphics[width=7.5cm,height=3.8cm]{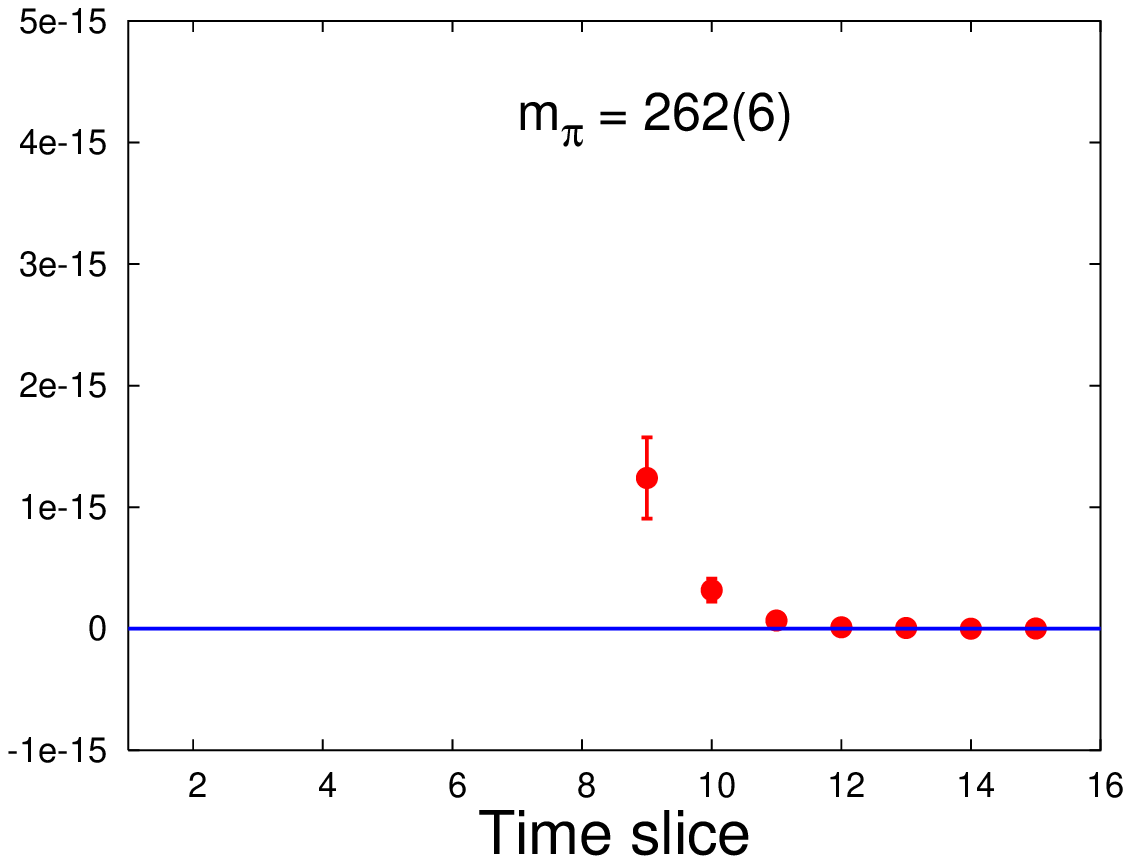}\\
\vspace*{-0.03in}
\includegraphics[width=7.5cm,height=3.8cm]{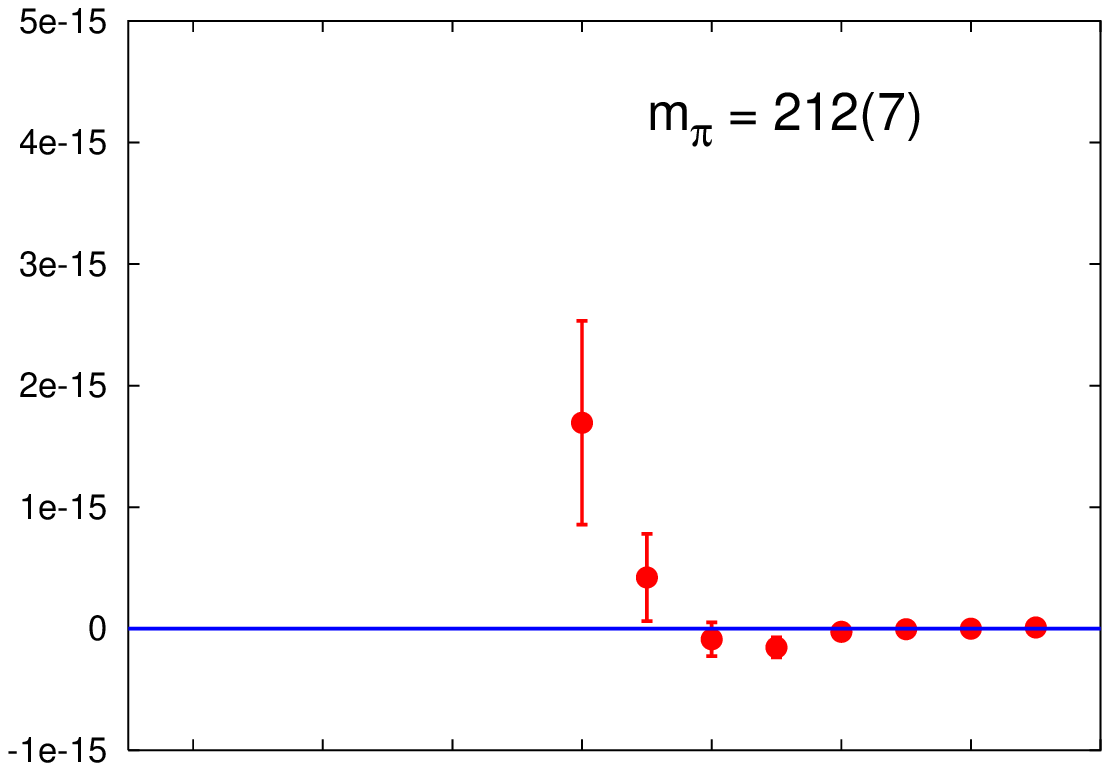}
\includegraphics[width=7.5cm,height=3.8cm]{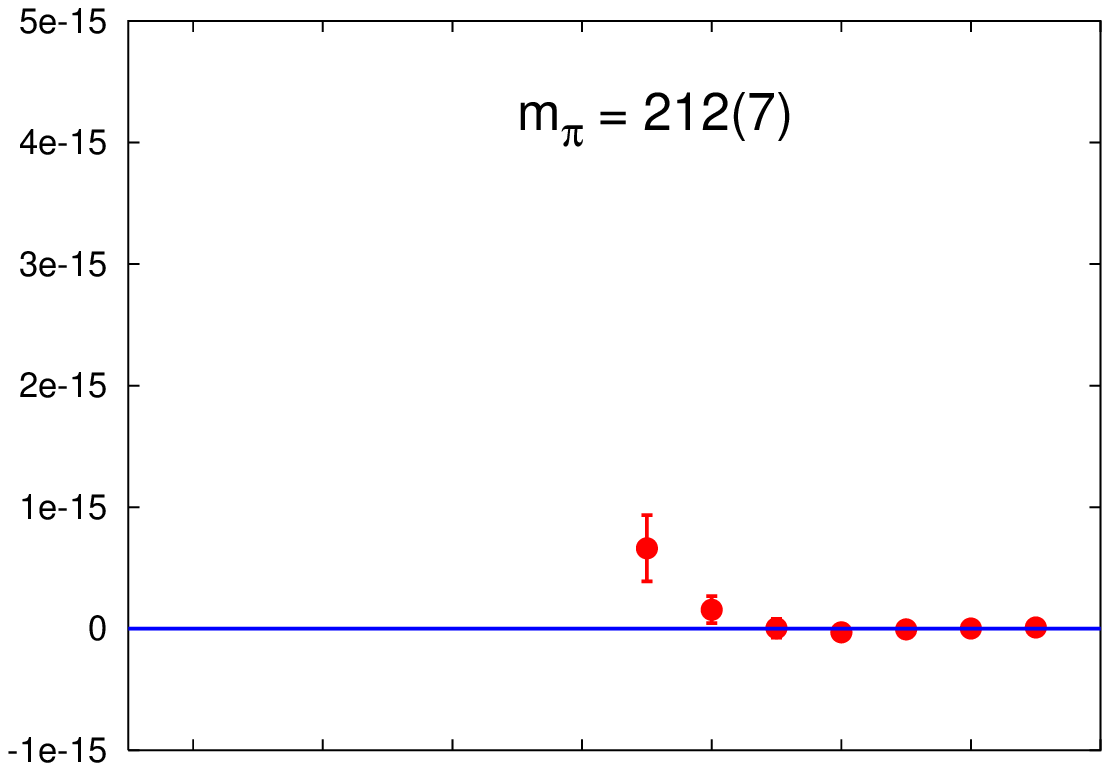}\\
\vspace*{-0.22in}
\includegraphics[width=7.5cm,height=3.8cm]{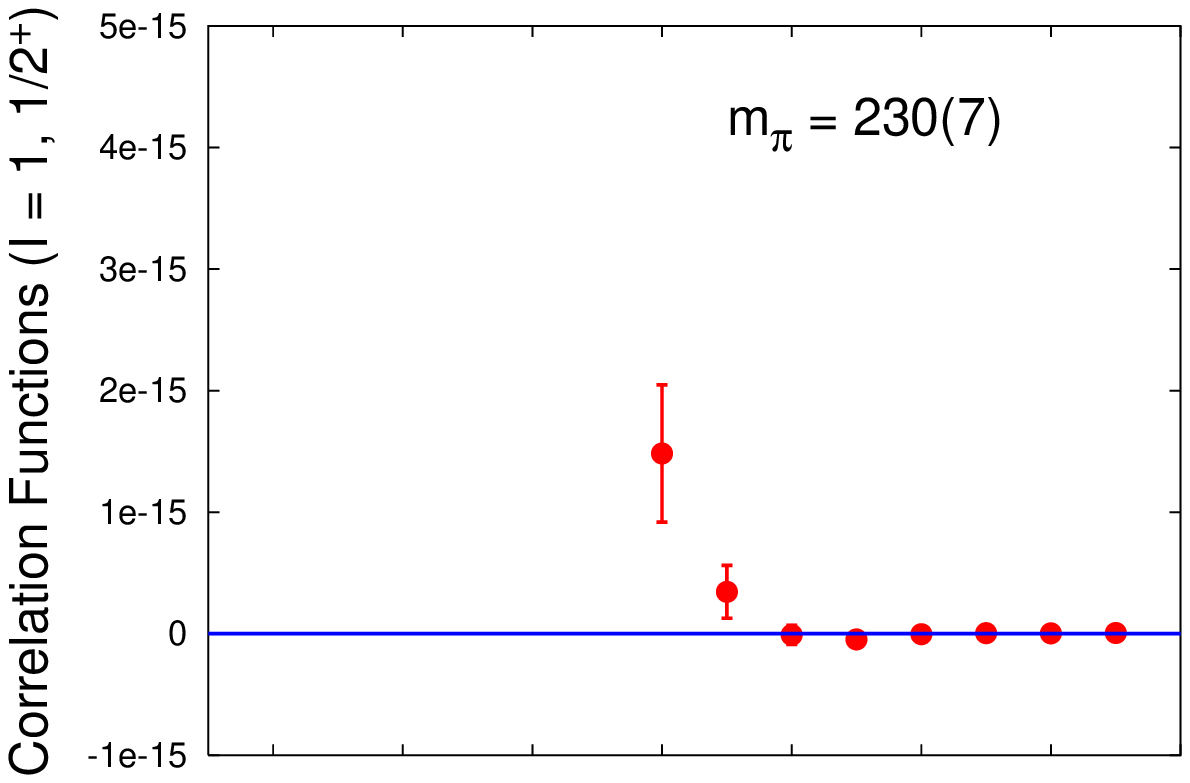}
\vspace*{-0.22in}
\includegraphics[width=7.5cm,height=3.8cm]{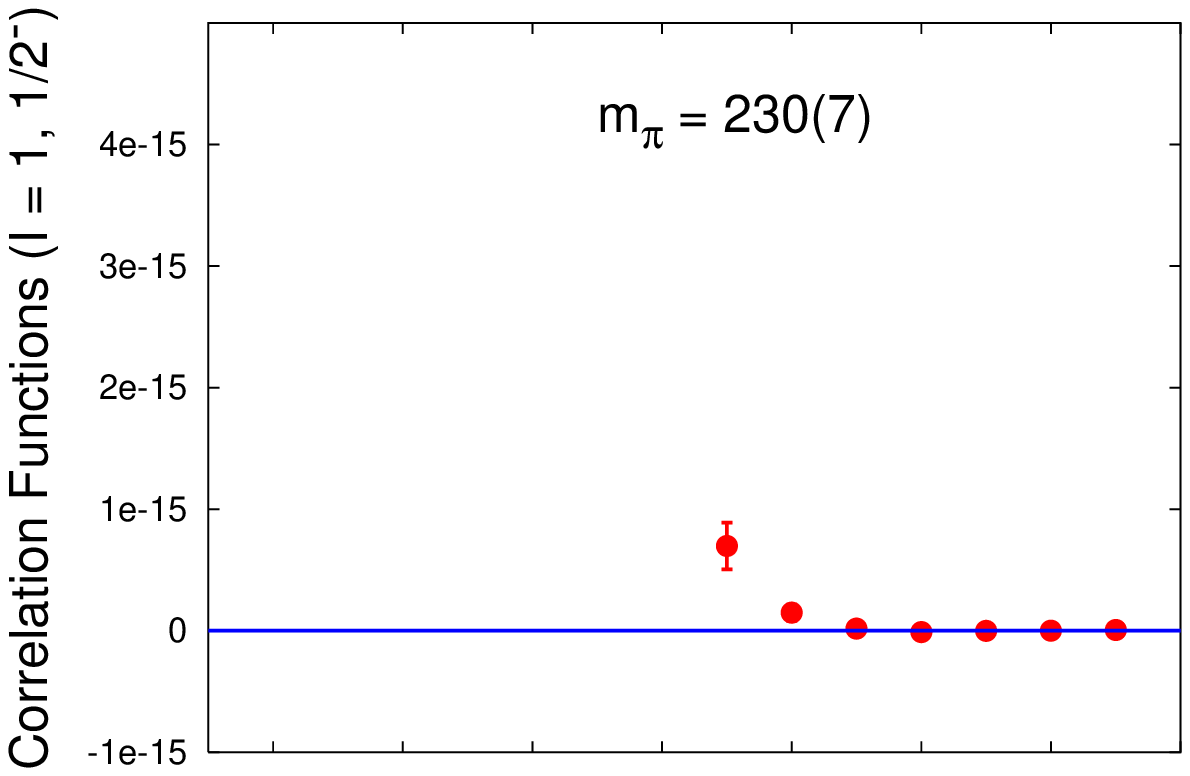}\\
\includegraphics[width=7.5cm,height=3.8cm]{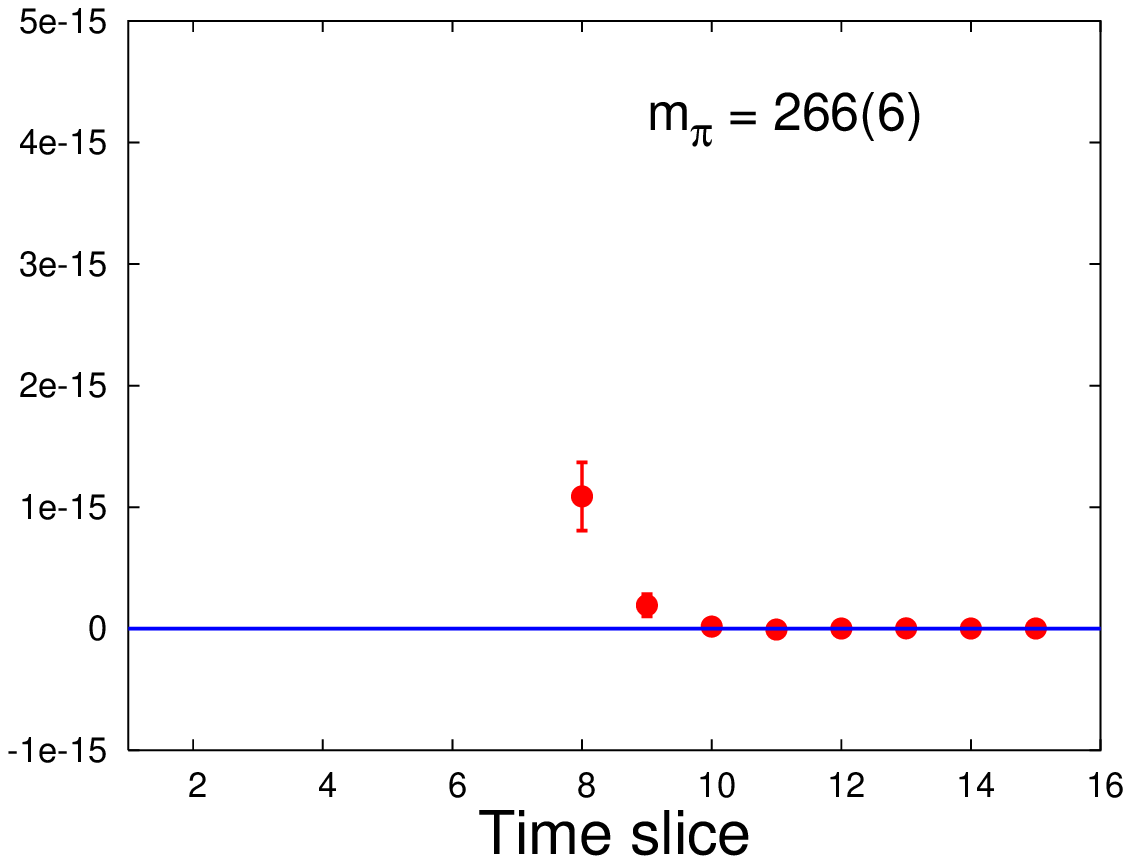}
\includegraphics[width=7.5cm,height=3.8cm]{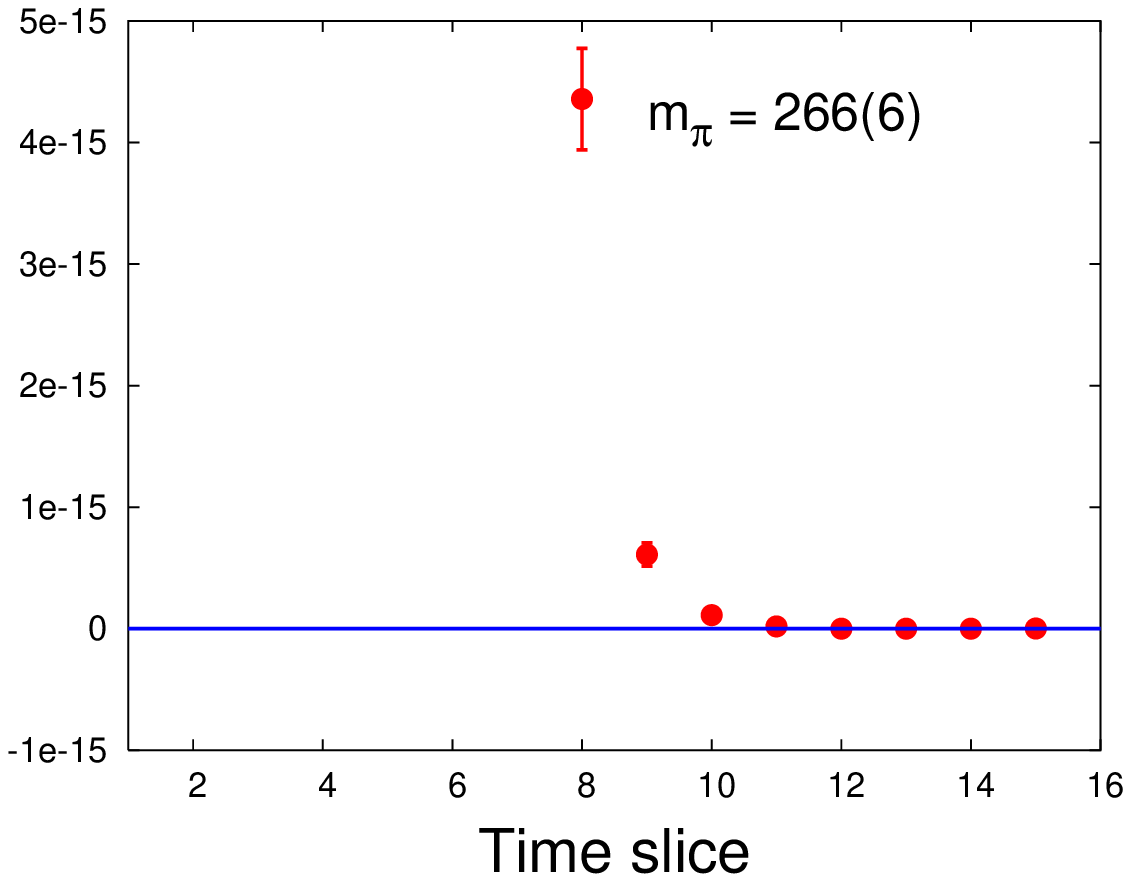}
\vspace*{-0.1in}
\caption{Correlation functions for the positive parity (first column) 
and the negative parity (second column) for 2.4 fm (first three rows) and 3.2
fm (second three rows) lattices, respectively.  These correlation
functions are for three pion
masses (in MeV). One should notice that there are negative dips in
the positive parity channel which indicate the presence of ghost
states. These $KN\eta^{\prime}$ ghost states decouple from the
correlation function at pion mass above $\sim$ 300 MeV.}
\label{cfun2}
\end{figure*}

\subsection{Ghost states}
Since we are working in the light quark region with the quenched
approximation, there is an additional complication arising from
hairpin diagrams corresponding to the would-be $\eta^\prime$ meson
which is degenerate in mass with the pion.  The resulting states
containing the would-be $\eta^\prime$ are called ghost states since
they are unphysical quenched artifacts with negative spectral weights.
Because of the negative weight the key signature for the presence of
ghost states is the negative correlation function at very small pion
mass.  Such quenched artifacts have been observed in both the
meson~\cite{bde02} and baryon~\cite{roper04} sectors.  In the case of
pentaquarks, the ghost state is $NK\eta^\prime$ which has intrinsic
positive-parity with relative $S$-waves among the hadrons.  In the
$1/2^-$ channel, there is a relative $P$-wave between a pair of $N$,
$K$, and the ghost $\eta'$. The threshold of this state is thus raised
on the lattice: for example,  
$E_{KN\eta^\prime} = E_{K} + E_{N} + E_{\eta^\prime} =\sqrt{m_K^2+p_{K}^2}\, 
+ m_N+\sqrt{m_{\pi}^2+p_{\pi}^2}$
where $p$ starts from $n=1$ for a $P$-wave between $K$ and $\eta'$. On
our smaller lattice, the value is about 2.23 GeV at 188 MeV pion mass.
So the ghost state in the $1/2^-$ channel lies quite high while the
lowest states are expected to be the KN scattering state in $S$-wave
threshold and the possible pentaquark.

In the $1/2^+$ channel, the situation is the opposite.  The $K$, $N$,
and the ghost $\eta'$ are all in relative $S$-wave which has a
threshold of $E_{KN\eta^\prime}(n=0)=m_K+m_N+m_\pi$. Using
experimental values for pion, nucleon and kaon
$E_{KN\eta^\prime}(n=0)=1.57$ GeV, which is right near the possible
pentaquark of 1.54 GeV.  So in the $1/2^+$ channel, the ghost state
lies relatively low and plays a significant role, just like in the
$S_{11}(1535)$ channel where the ghost state lies lower than the
$S_{11}$ when the quark mass is light~\cite{roper04} (i.e.with pion
mass lower than 300 MeV).  So in this channel, the ghost state and the
potential pentaquark state are candidates for the ground state,
followed by the excited state of $n=1$ $KN$ scattering state. In our
algorithm, the ghost state is modeled as $-W_g(1+E_\pi t)e^{-E_gt}$
where $E_{\pi}=\sqrt{m_\pi^2+p_{\pi}^2}$ and $E_g$ is constrained near its
threshold value~\cite{roper04}. The fact that the ghost state has a
negative spectral weight $-W_g$ is crucial for its isolation from the
non-ghost states.

In Fig.~\ref{cfun1}, we present I = 1 correlation functions (at a pion mass
around 200 MeV) for both parity channels and for two lattice volumes
(2.4 and 3.2 fm). The top two figures are for 2.4 fm and bottom two
are for 3.2 fm. A row-wise comparison of these figures reveals that
the positive parity correlation functions have a negative dip which
indicates the presence of quenched ghost states. On the other hand,
correlation functions for the negative parity channel are always
positive which confirms the absence of low-lying ghost states in that
channel.  Comparing the left two figures (one also has to notice the
difference of scales) it is clear that the ghost states in the smaller
volume are more prominent.

In Fig.~\ref{cfun2}, we plot a few more I = 1 correlation functions.  As in
Fig.~\ref{cfun1}, the left side figures are for positive parity and
the top six figures are for the smaller volume lattice.  Figures on the
left side show the effect of ghost states in the correlation function.
Note that the ghost states are only noticeable in the very light quark
mass region.  The effect of the ghost states decreases as the pion
mass increases, and decouples from the correlation function near pion
mass around 300 MeV.  The right side figures are for the negative
parity channel and they are always positive. In the previous lattice
calculations~\cite{csikor,sasaki,chiu}, the pion masses are above 440
MeV, and therefore, there is no need to consider the ghost
states. Since our lowest pion mass is 182 MeV, we will need to take
the ghost states into account in the $1/2^+$ channel.

\begin{figure}
\includegraphics[height=6.0cm]{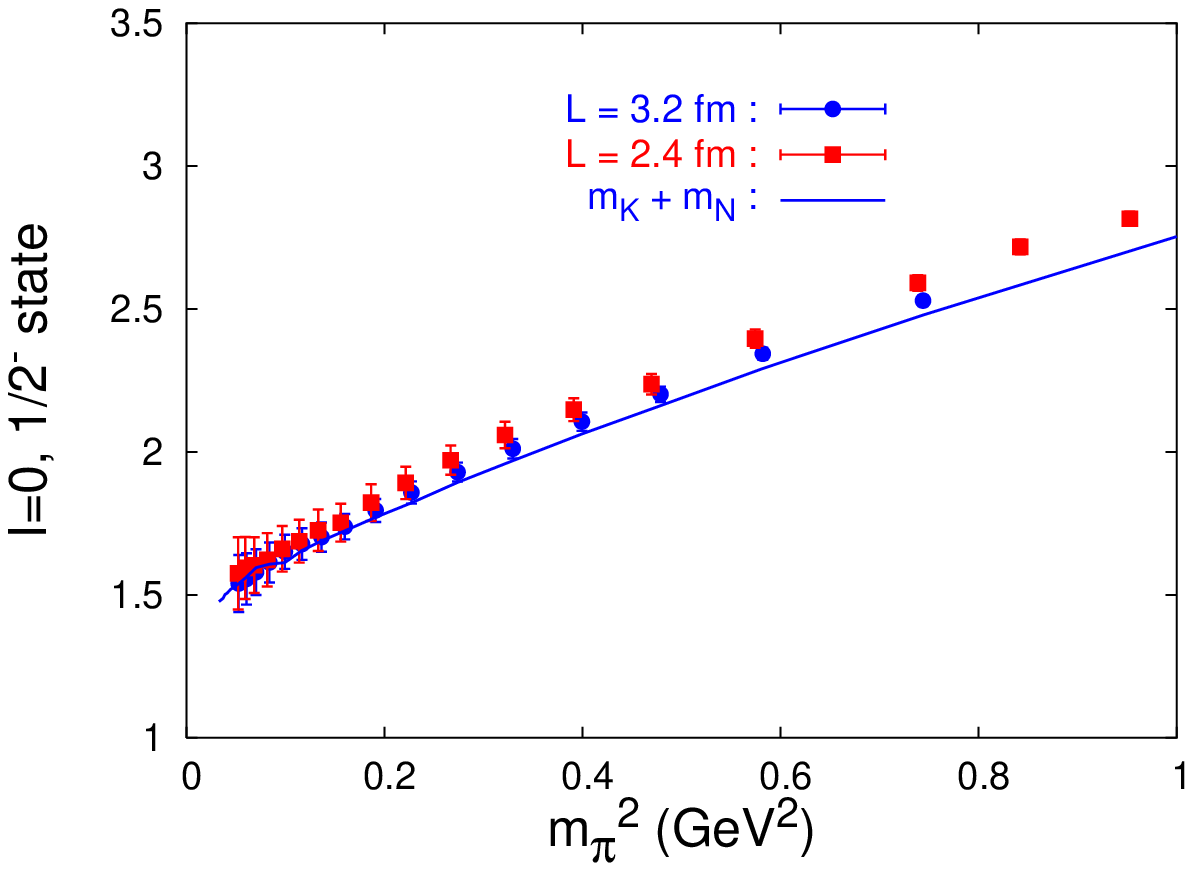}
\includegraphics[height=6.0cm]{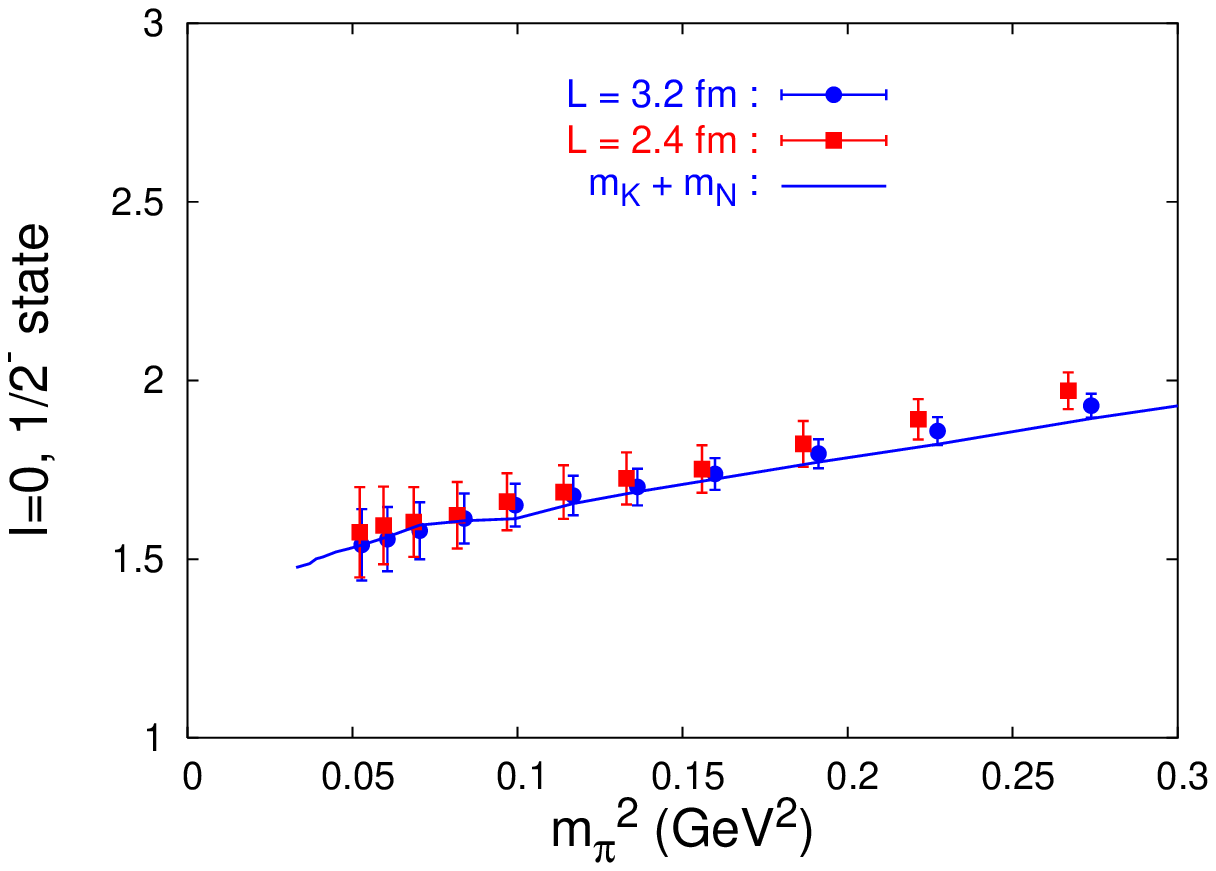}
\caption{The computed ground state mass in the
$I(J^P)=0\left(1/2^-\right)$ channel
as a function of $m^2_\pi$ for the two lattices L=2.4 fm and L=3.2 fm.
The curve is the KN threshold energy in the $S$-wave
$E_{KN}(n=0)=m_K+m_N$. The bottom  figure is an enlarged version of
the top figure for the small quark mass region.}
\label{I0_m}
\end{figure}

\begin{figure*}
\includegraphics[height=6.0cm]{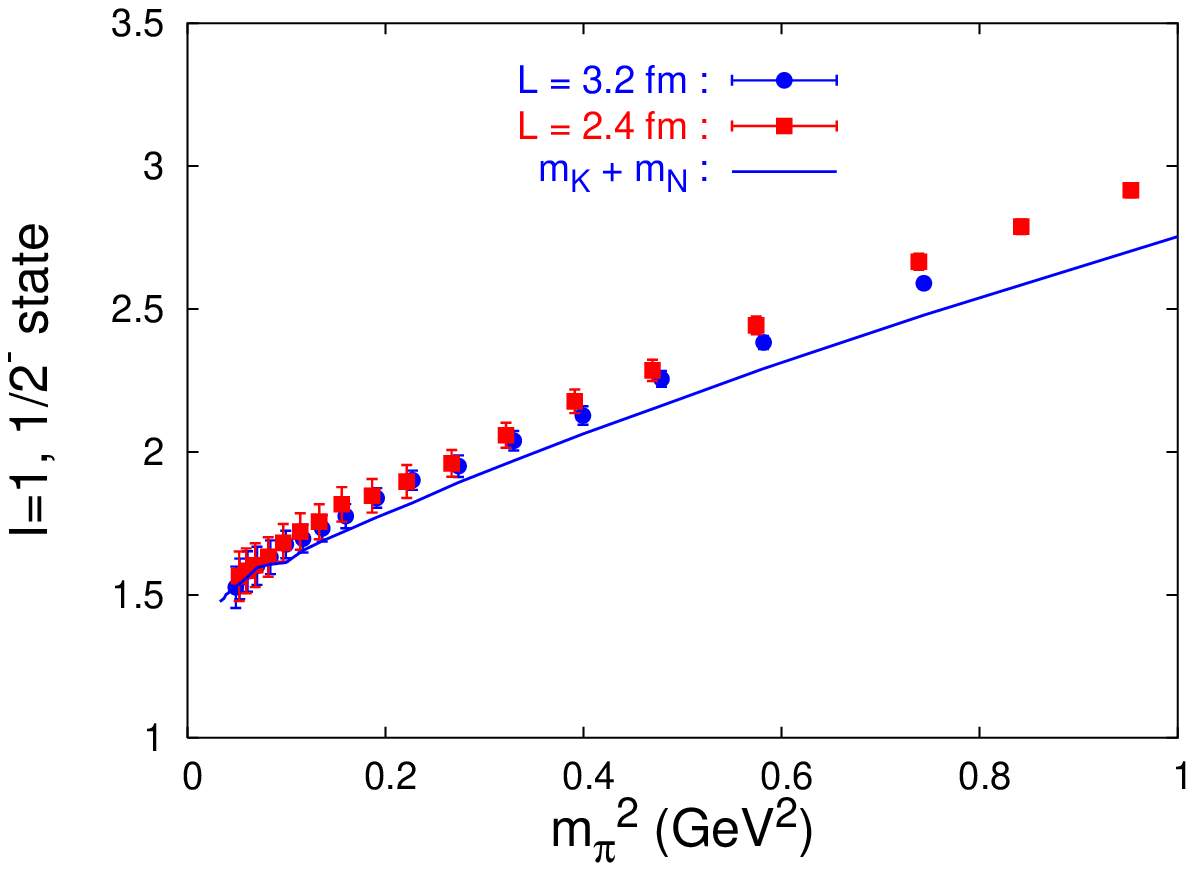}
\includegraphics[height=6.0cm]{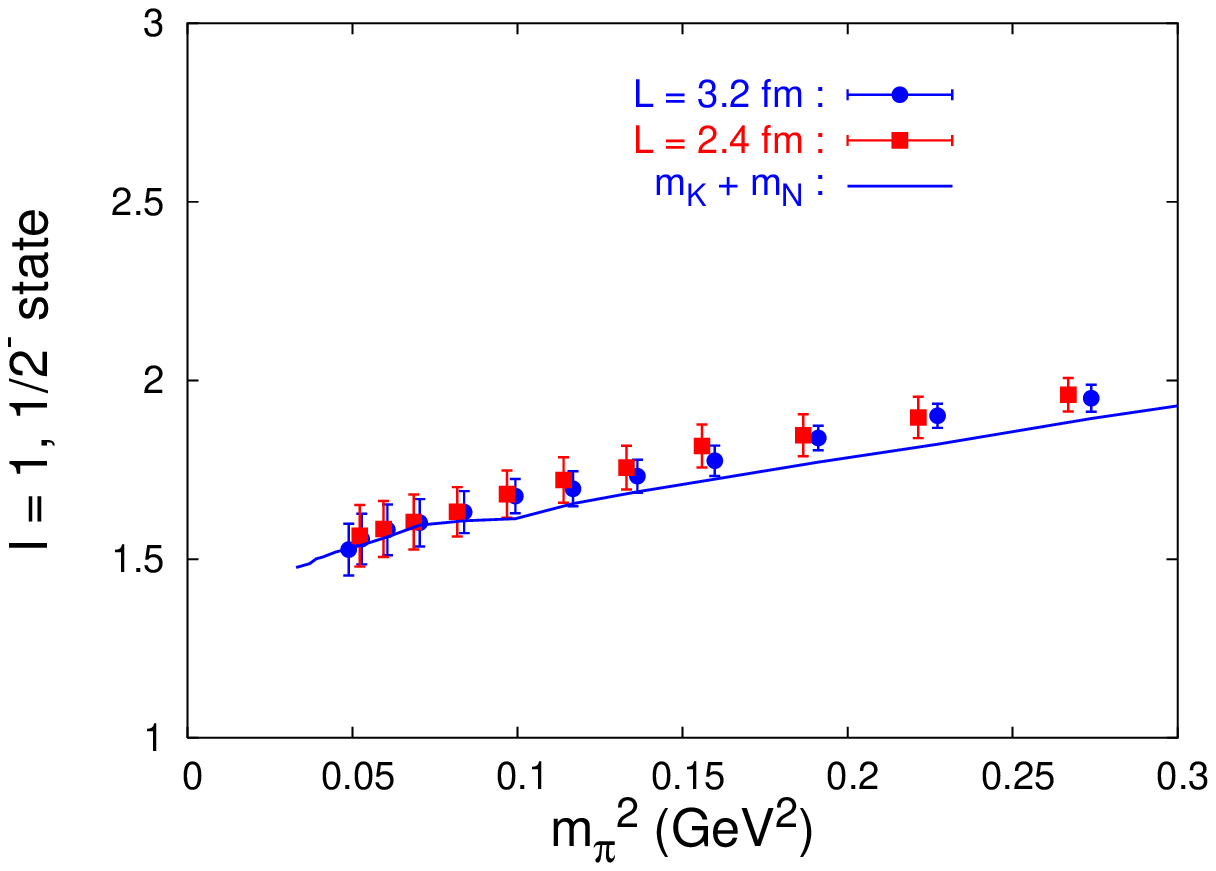}
\caption{The computed ground state mass in the
$I(J^P)=1\left(1/2^-\right)$
channel as a function of $m^2_\pi$.
As above, the curve is the KN threshold energy in the $S$-wave
$E_{KN}(n=0)=m_K+m_N$, and the bottom figure is an enlarged version of
the top figure for the small quark mass region.}
\label{I1_m}
\end{figure*}

\begin{figure*}
\includegraphics[height=6.0cm]{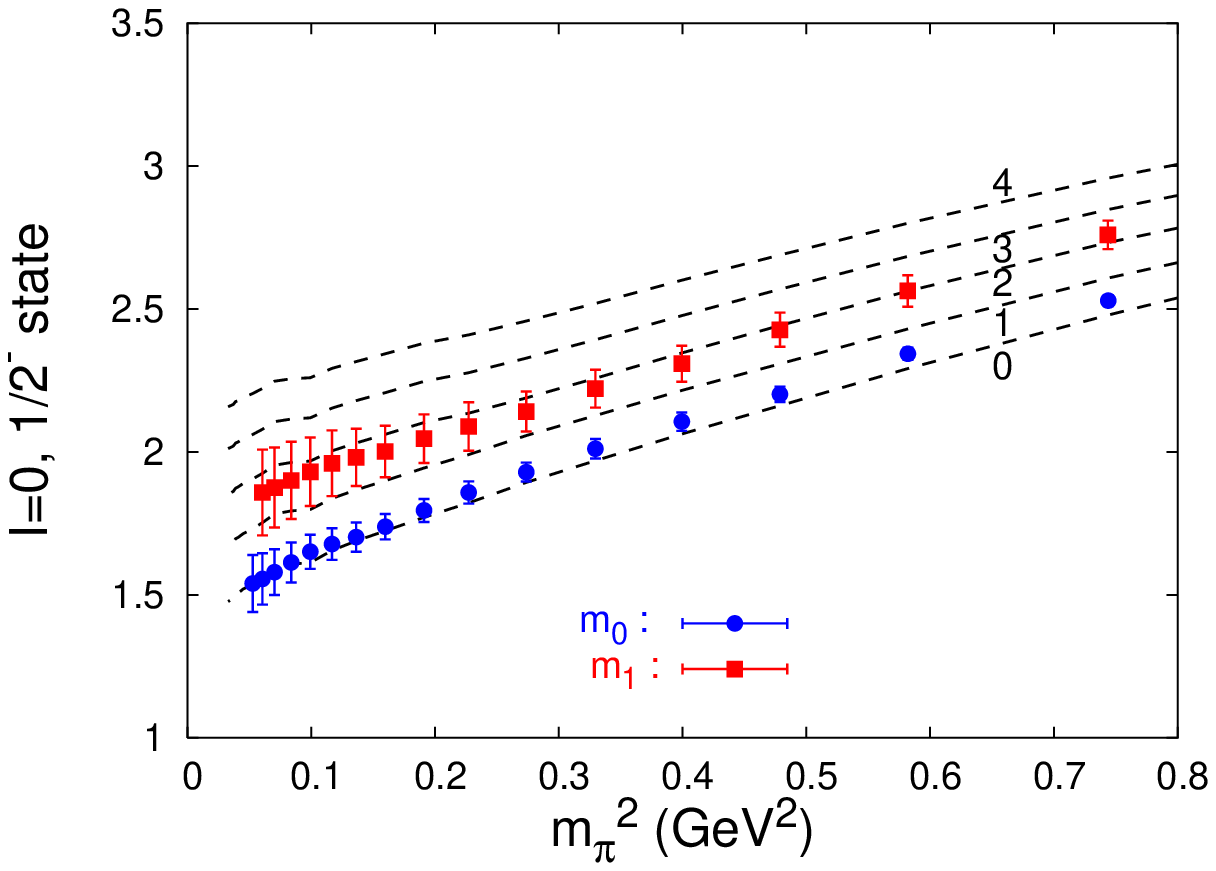}
\includegraphics[height=6.0cm]{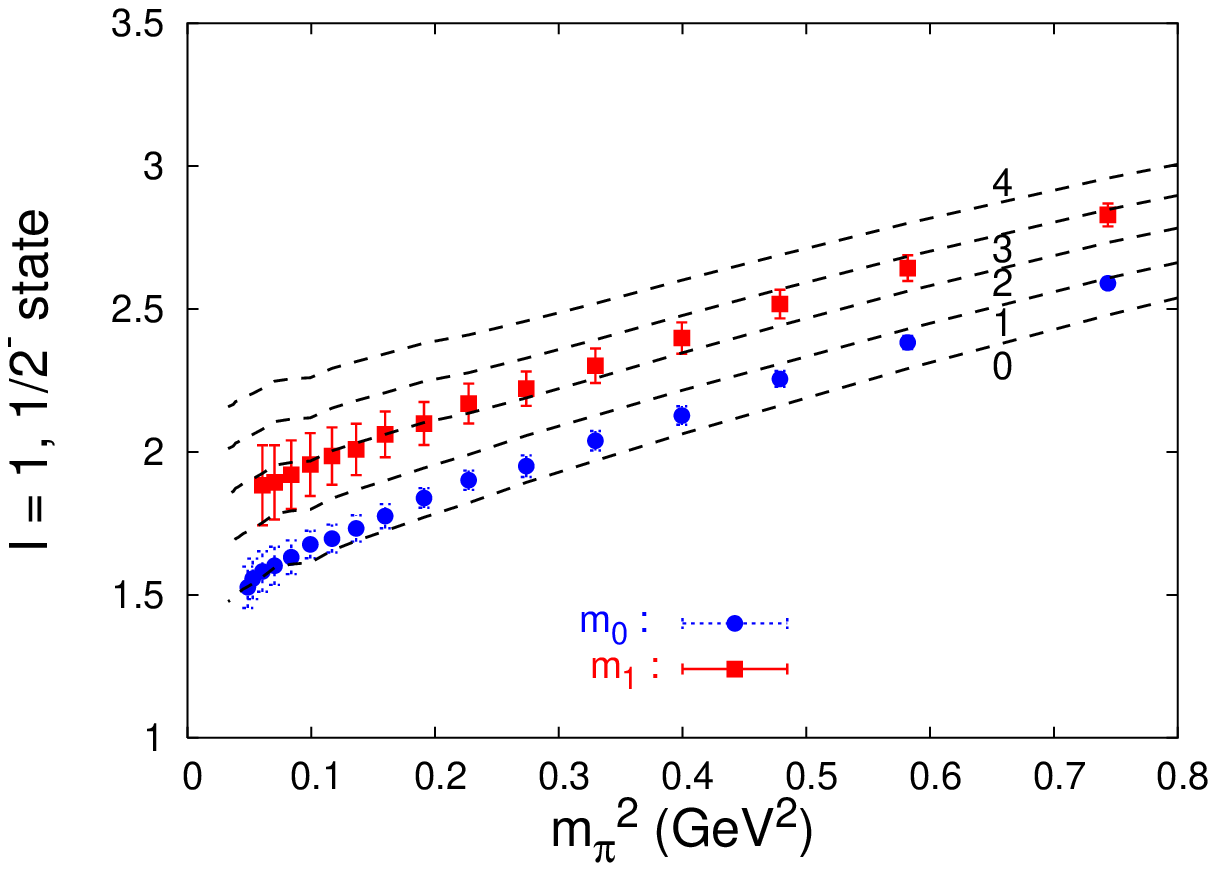}
\caption{The computed masses in the $I(J^P)=0\left(1/2^-\right)$ (top) $1\left(1/2^-\right)$ (bottom) 
channels as a function of $m^2_\pi$ (for 3.2 fm lattice). The first
excited states (square) are shown along with the ground state (circle).
The first few scattering states corresponding to different discrete
momenta (for $n$ = 0,1,2,3 and 4) are shown by dashed lines.}
\label{I1_m_excited}
\end{figure*}


\subsection{Negative-parity channel}
In Figs.~\ref{I0_m} and~\ref{I1_m} we show the results for 
the ground state mass as a function of $m_\pi^2$ in the
$I(J^P)=0\left(1/2^-\right)$ and  $1\left(1/2^-\right)$ channels
respectively. Also plotted is the KN threshold energy in the $S$-wave
$E_{KN} (n=0) = m_K+m_N$ which is the same on both lattices.  There is no
need to consider ghost states in these channels, which is supported by
the fact that the correlation functions are positive throughout.  The
ground state energy on the smaller lattice (L=2.4 fm) is consistently
higher than that on the larger one (L=3.2 fm).  At the lowest mass,
the energy coincides with the $S$-wave threshold, meaning that there
is little interaction, which is consistent with the experimental fact
of zero scattering length (see Table~\ref{tab_len}).

Here we point out that the uncertainty in the strange mass
determination (which we set by the $\phi$ meson and resulting in our
kaon mass being higher than the experimental value by about 7\%) is
harmless. We note that changing the kaon mass affects both the data
and the energy threshold simultaneously, so that the difference
between them is essentially unaffected.

We notice that at larger $m_{\pi}^2$, the fitted ground state is
higher than the $m_K + m_N$ threshold.  In the higher quark mass
region, the excited states have larger weight relative to the ground
state.  This, together with a somewhat smaller mass gap between
different momentum states at higher quark masses, results in an
effective mass which continues to drop at the largest time slices. As
a result, in the higher quark mass region it is more difficult to
isolate the ground state, and the fitted lowest state is a mixture of
the ground state and the excited states with $n = 1$ and higher. The
fact that the fitted lowest state energy on the smaller lattice (L=2.4
fm) is higher than that on the larger one (L=3.2 fm) for the same
higher $m_{\pi}^2$ is consistent with this interpretation since the
lattice momenta on the smaller lattice is larger than that on the larger
lattice, leading to a mixed state with higher energy in the smaller
lattice. In the low quark mass region, the excited states have smaller
relative weights, and thus we are able to fit the ground state more
accurately close to the mass threshold (i.e. $m_K + m_N$ with $n =
0$). However, the lack of statistics prevented us from resolving the
$n = 1, 2 $ and 3 states which are closely packed
together~\cite{fitting04}. As can be seen from Table~\ref{tab_KN}, the
separation of KN states with different momenta ranges from 150 to 200 MeV
which are much lower than the usual separation between the ground
states and their radial excitations in ordinary hadrons. The latter
are usually about 500 - 600 MeV. Consequently, our fitted first
excited states appear to be a mixture of the $n =1$ and higher momenta
states.

\begin{figure}
\includegraphics[height=6.0cm]{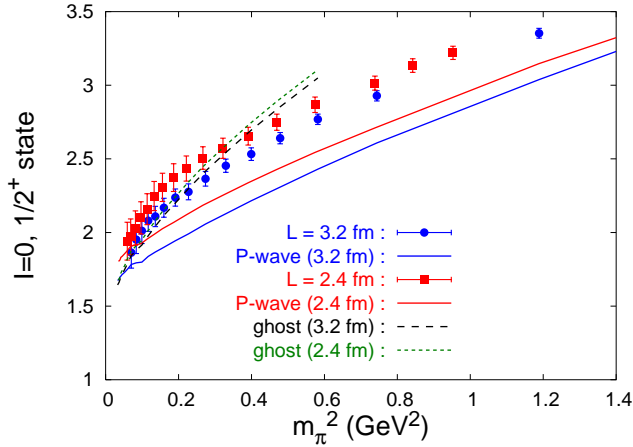}
\caption{The computed mass in the $I(J^P)=0\left(1/2^+\right)$ channel
as a function of $m^2_\pi$ for the two lattices L=2.4 fm and L=3.2 fm.
The two lower curves are the KN threshold energies in the $P$-wave
$E_{KN}(n=1)$.  The two higher curves represent the energies of the
non-interacting ghost states.}
\label{I0_p}
\end{figure}
\begin{figure}
\includegraphics[height=6.0cm]{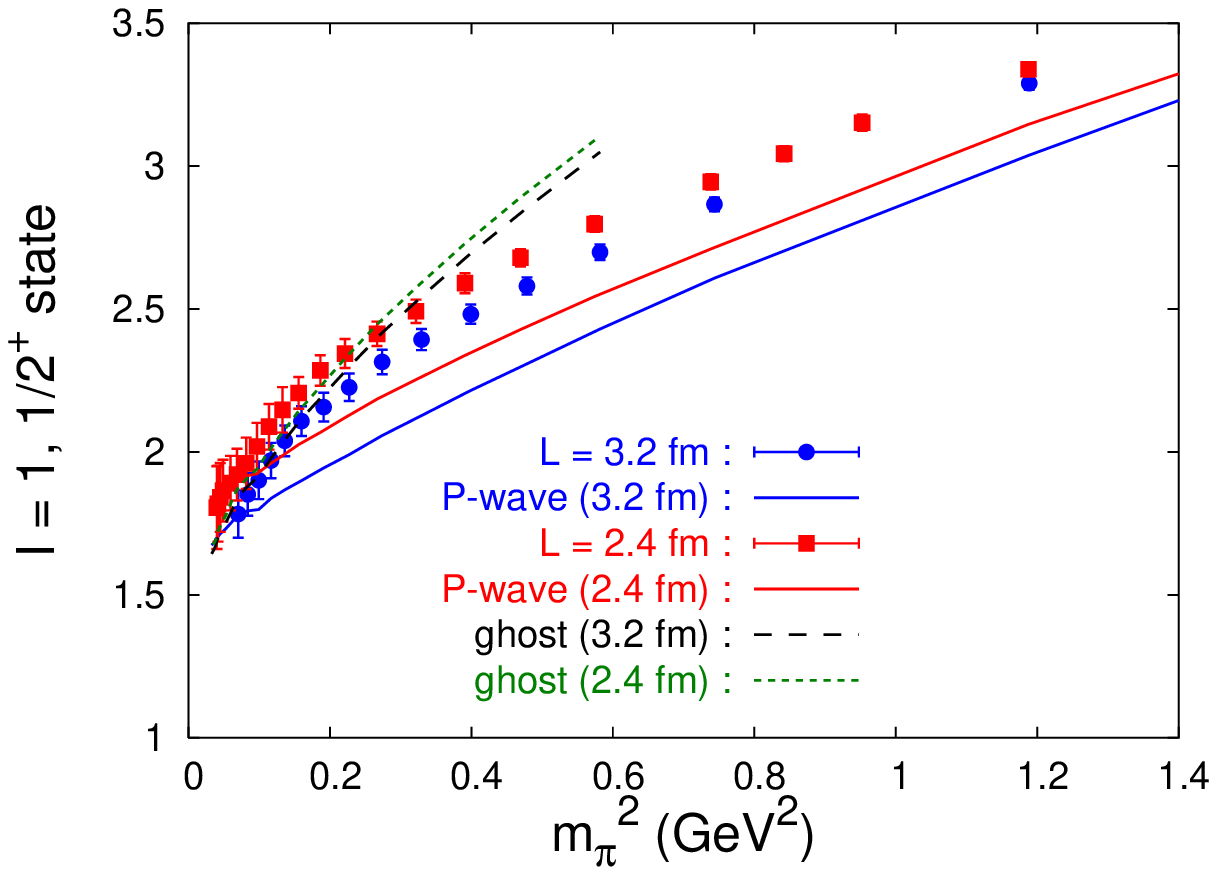}
\includegraphics[height=6.0cm]{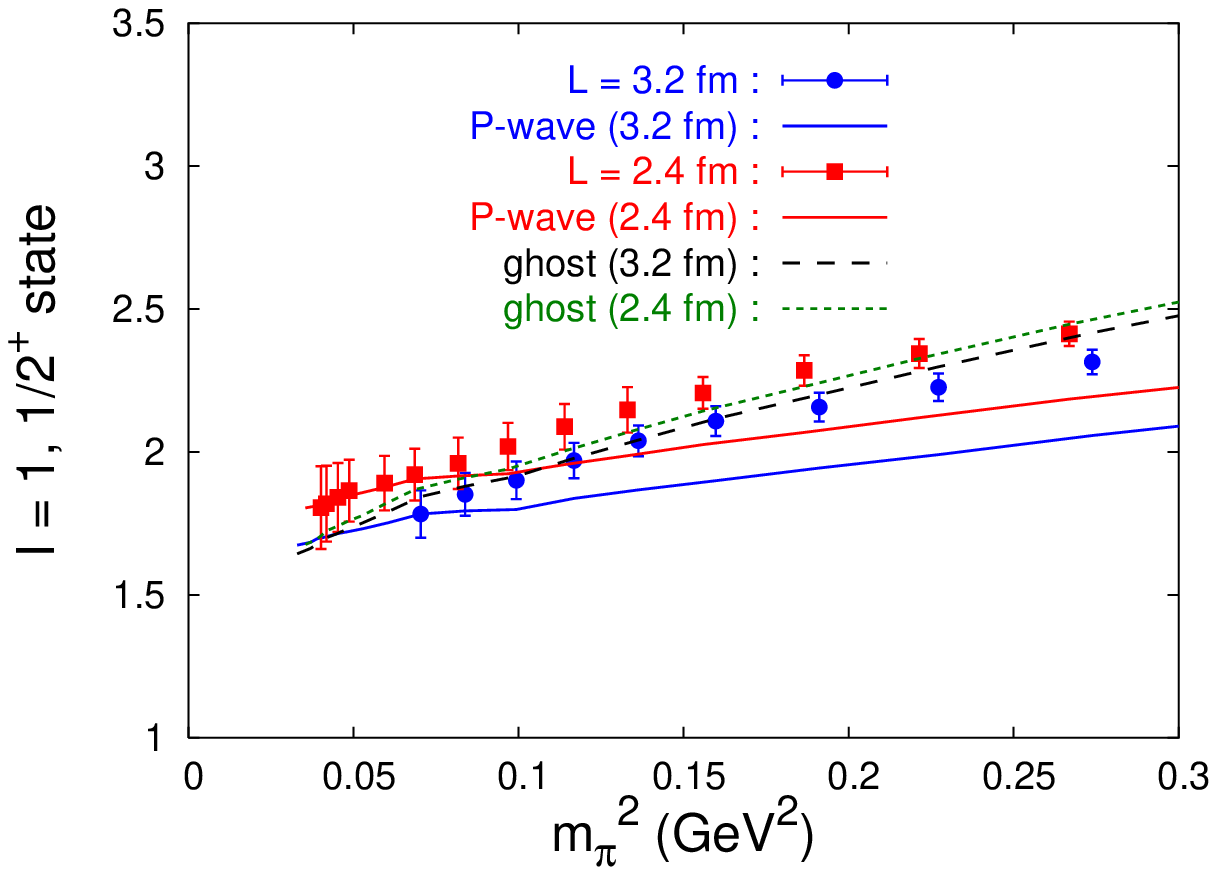}
\caption{The computed mass in the $I(J^P)=1\left(1/2^+\right)$ channel
as a function of $m^2_\pi$ for the two lattices L=2.4 fm and L=3.2 fm.
The two lower curves are the KN threshold energies in the $P$-wave
$E_{KN}(n=1)$.  The two higher curves represent the energies of the
non-interacting ghost states. The bottom figure is an enlarged version
of the top figure in the lower quark mass region.}
\label{I1_p}
\end{figure}

As far as the ground states are concerned, our results more or less
agree with those of Ref.~\cite{csikor} and \cite{sasaki}, but disagree
qualitatively with those of Ref.~\cite{chiu}. It is noted in
Ref.~\cite{csikor} and \cite{sasaki} that they have seen a low-lying
excited state above the $KN$ mass threshold and they interpret it as
the pentaquark state. From the top figure of 
Fig.~\ref{I1_m_excited}, we see that the
first excited state near the chiral limit is $\sim 1.85$ GeV which is
substantially higher than 1.54 GeV, the experimentally observed
pentaquark state. As explained above, we interpret it as the mixture
of the $n=1$, $n=2$, and possibly $n=3$ KN scattering states.  We
observe similar behavior for the I = 1 case also (bottom figure). There
is no candidate for a pentaquark between the KN threshold ($n= 0$) 
and the first KN scattering state ($n =1$) which is contrary to the finding of
Ref.~\cite{sasaki}. We tried to accommodate an extra low-lying
pentaquark state in between our ground and first excited states.
However, the $\chi^2$ fit always rejects such an intermediate
state. This implies that our data do not favor a pentaquark state in
between the two scattering states with lowest momenta.  This, however,
does not preclude a pentaquark state nearly degenerate (i.e. within
100 MeV) with the KN scattering state at threshold as was observed in
Ref.~\cite{csikor} with a two-channel approach. We shall study this
possibility in the future~\cite{penta2}.

\subsection{Positive-parity channel}
The fitted ground state mass as a function of $m^2_\pi$ in the 
$I(J^P)=0\left(1/2^+\right)$ and $1\left(1/2^+\right)$ channels are shown
in Fig.~\ref{I0_p} and  Fig.~\ref{I1_p} respectively. 
In these channels, the ghost $NK\eta^\prime$ must be included, 
as discussed above.  In the
fitting model, the $NK\eta^\prime$ ghost state, pentaquark, and KN
$P$-wave scattering state are considered.  We found a ghost state and
a KN scattering state, but not a pentaquark state near 1.54 GeV. In the
lower energy I = 1 channel, we have tried to 
see if our data could accommodate three states, but the
$\chi^2/dof$ would simply reject it.  The energy of the $KN$
scattering state lies higher on the smaller lattice (L=2.4 fm) than
that on the larger lattice (L = 3.2 fm). This mainly reflects the fact
that the momentum corresponding to $n=1$ is larger on the L = 2.4 fm lattice 
than that of on the L =
3.2 fm lattice.  At the lowest mass, the energies almost coincide with
the $P$-wave thresholds, meaning that the $KN$ interaction is weak,
consistent with experiment.  At higher $m_{\pi}^2$, we again observe
that the ground state lies substantially higher than the $P$-wave KN
threshold energy. We believe this is a result of a mixture with
several higher momenta states, as proposed in the last section to explain
the similar behavior for the negative parity states.
\begin{figure}[htb!]
\includegraphics[height=6.0cm]{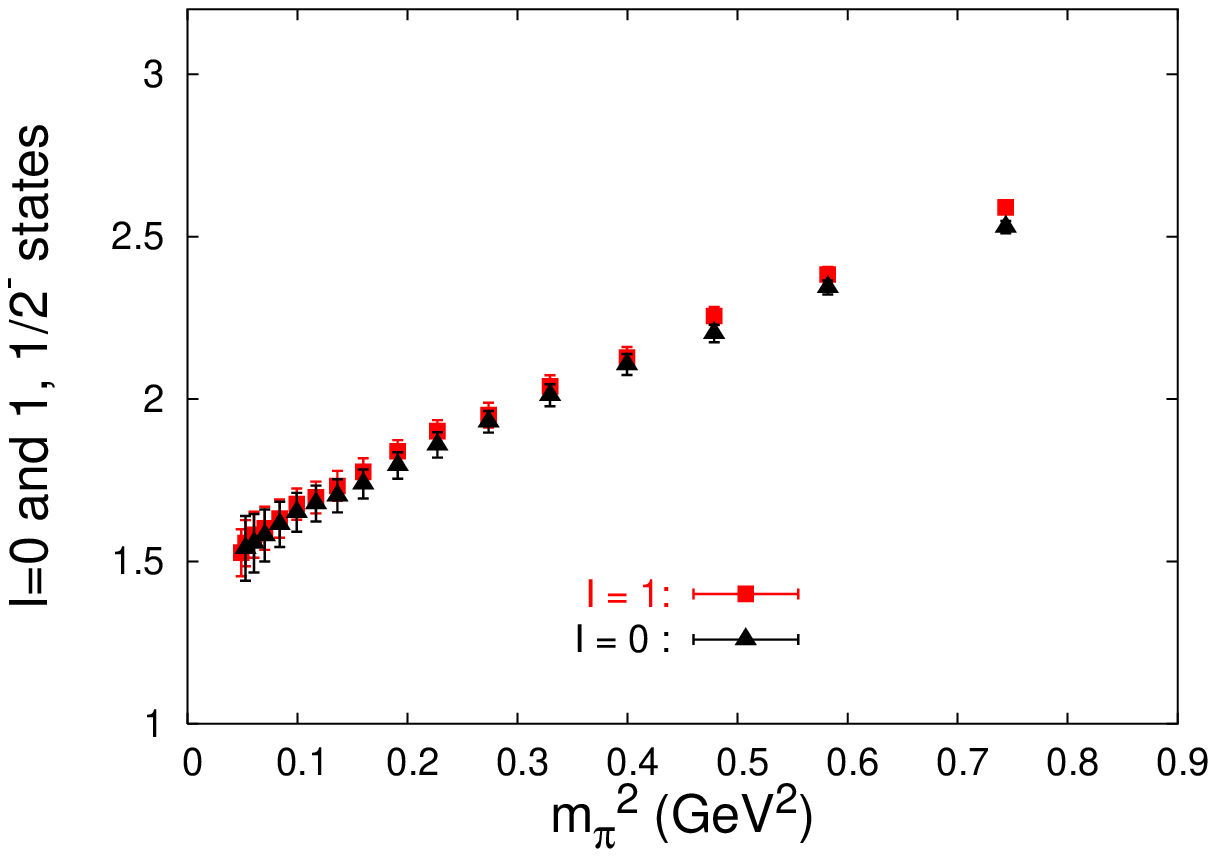}
\includegraphics[height=6.0cm]{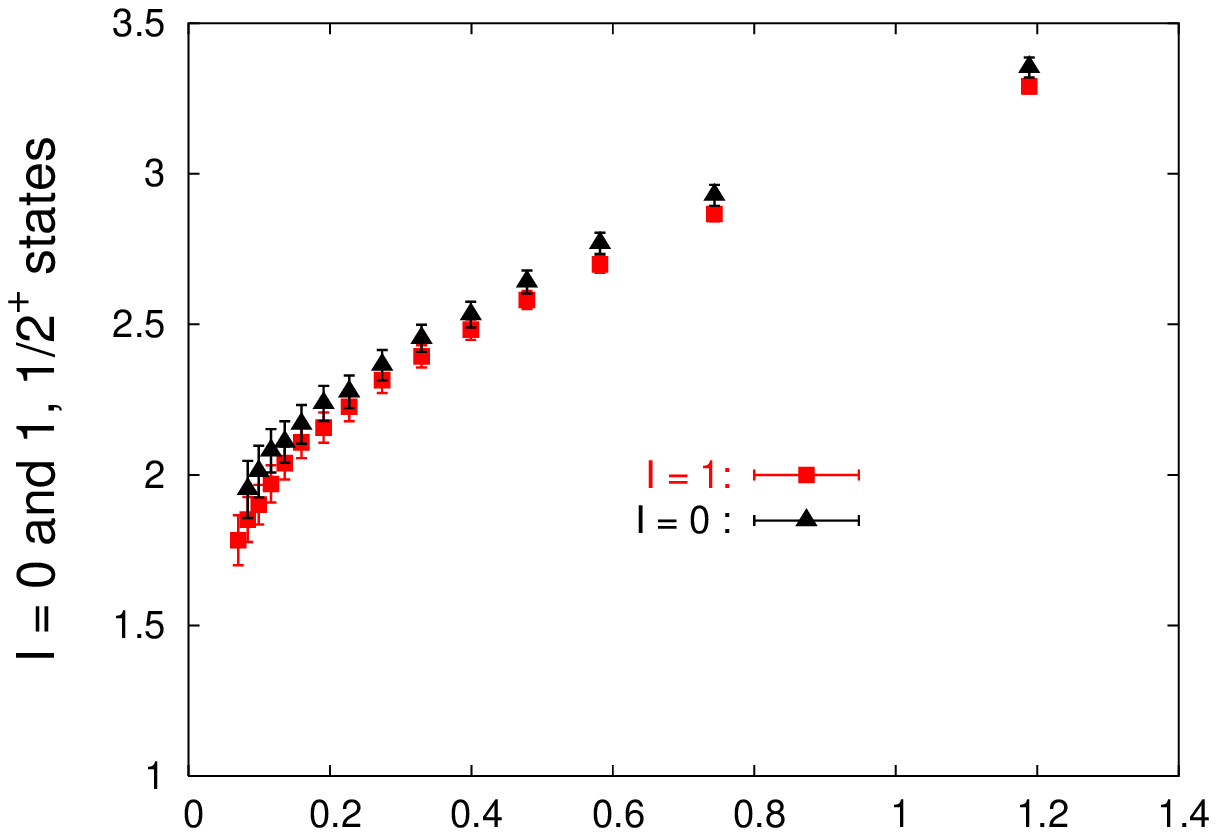}
\caption{A comparison of ground state masses as a function of $m^2_\pi$
for I = 0 and I = 1 channels both for the negative (top) and the positive
(bottom) parities.}
\label{I0_p_I1}
\end{figure}

In Fig.~\ref{I0_p_I1}, we compare the ground state energies for I=0
and I=1 on the 3.2 fm lattice. We observe that in the ${1\over 2}^{-}$
channel, although within
errorbars, the energy of the I=1
state tends to be higher for all masses. This tendency is inverted for
the ${1\over 2}^{+}$ channel. 
Both these observations are consistent with Ref.~\cite{csikor}.
Our conclusion that the pentaquark state is absent below the KN $P$-wave
threshold again agrees with those of Ref.~\cite{csikor,sasaki} 
and disagrees with that of Ref.~\cite{chiu}.

\subsection{Volume dependence}
In a box, lattice states have a volume dependence of $\sqrt{1/V}$ (where $V$
is the spatial volume) from the normalization on the lattice.  For a
one-particle state, a point-source and zero momentum point-sink 
correlation function is given by
\begin{eqnarray}
G(t)&=&\sum_{\vec{x}}\langle0|\chi(\vec{x},t)\bar{\chi}(0)|0\rangle
\nonumber\\
&=&\sum_{n} V {{|\langle0|\chi(0)|n\rangle|^{2}}\over {2M_{n} V}}\, e^{-M_{n}t}
\nonumber\\
&=&\sum_{n}  W_{n} e^{-M_{n}t}
\label{w1}
\end{eqnarray}
where 
\begin{eqnarray}
W_{n} = {{|\langle0|\chi(0)|n\rangle|^{2}}\over {2M_{n}}}
\end{eqnarray}
is the spectral weight corresponding to the mass $M_{n}$, which has 
no explicit volume dependence.
On the other hand, for a non-interacting two-particle scattering state, 
with a total zero momentum, the corresponding correlation function
can be written as
\begin{eqnarray}
G(t)&=&\sum_{\vec{x}}\langle0|\chi_{1}(\vec{x},t)\chi_{2}(\vec{x},t)\bar{\chi}_{1}(0)\bar{\chi}_{2}(0)|0\rangle \nonumber\\
&=&\sum_{n_{1}n_{2}} V {{|\langle0|\chi_{1}(0)|n_{1}\rangle|^{2}
|\langle0|\chi_{2}(0)|n_{2}\rangle|^{2}}\over {2E_{n_{1}}V\,2E_{n_{2}} V}}\, e^{-(E_{n_{1}}+E_{n_{2}})t} \nonumber\\
&=&\sum_{n_{1}n_{2}} {W_{n_{1}} W_{n_{2}} \over V}\, e^{-(E_{n_{1}}+E_{n_{2}})t}
\label{w2}
\end{eqnarray}
which has an explicit inverse volume factor. Though the total momentum
between the two particles is zero they can have finite lattice momentum,
and thus, in Eq.~\ref{w2} we used energies $E_{n_{1}}$ and $E_{n_{2}}$
instead of masses $M_{n_{1}}$ and $M_{n_{2}}$. Since the $KN$
interaction is very weak, approximating the spectral weight of the
$KN$ scattering state with the non-interacting expression
(Eq.~\ref{w2}) should be a reasonable one, as far as the volume
dependence is concerned. Besides the explicit volume dependence, the
spectral weight in the noninteracting case i.e. ${W_{n_{1}} W_{n_{2}}
\over V}$ has implicit volume dependence from $W_{n_{1}}$ and
$W_{n_{2}}$. However, in our case, the spectral weights for the
nucleon, Roper, and $S_{11}$ have very small volume dependence (at a
few percent level) between the $12^{3} \times 28$ and $16^{3} \times
28$ lattices~\cite{roper04}.  As a result, we can expect that even for
a worst case the combined volume factor from $W_{n_{1}}$ and
$W_{n_{2}}$ would be much smaller than the explicit volume factor $V$,
which is 2.37 for our lattices. Therefore, it can be concluded that
the spectral weight $W$ for a one-particle state should not exhibit
explicit volume dependence whereas a two-particle scattering state
will show an explicit volume dependence in its spectral weight.
\begin{figure}[htb!]
\includegraphics[height=6.0cm]{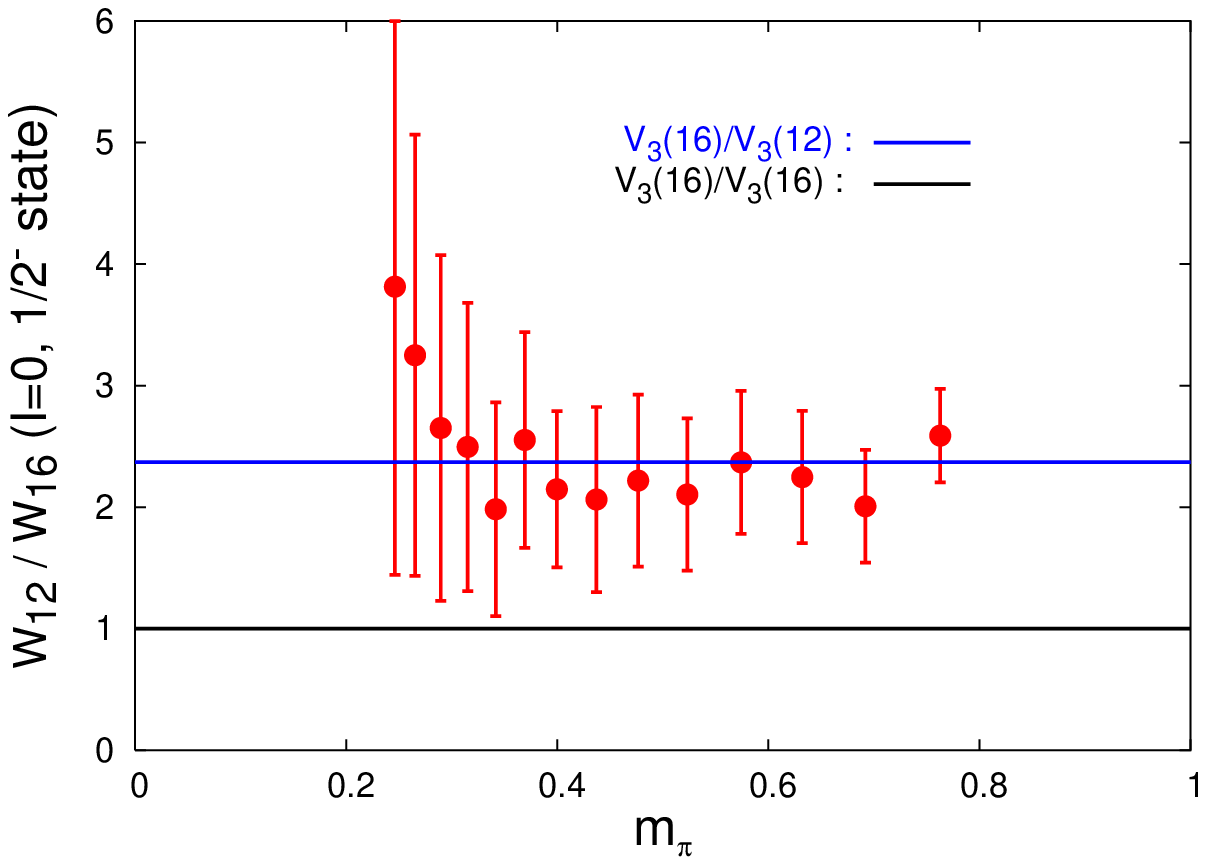}
\includegraphics[height=6.0cm]{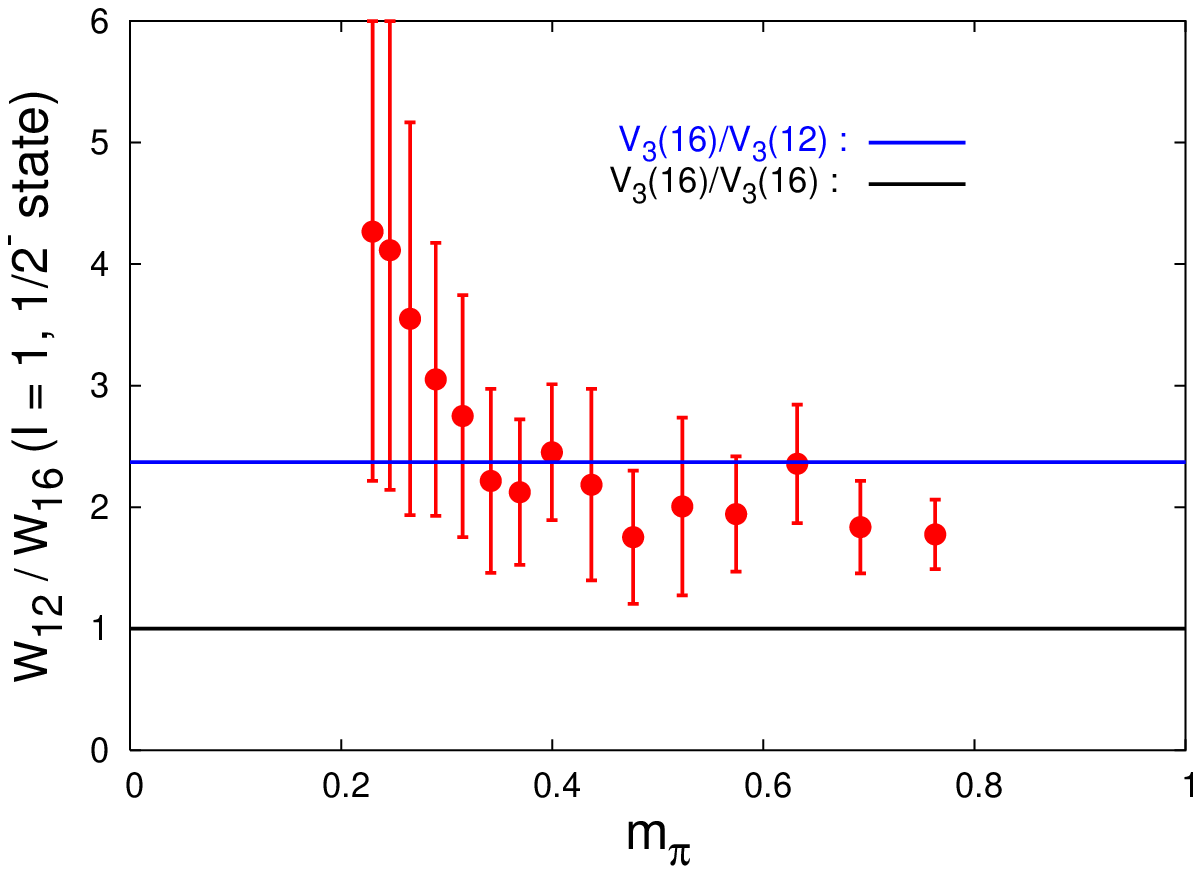}
\caption{Ratio of spectral weights on the two lattices in the negative parity 
 $I = 0$ (top figure) and $I = 1$ (bottom figure) channels. The line at 2.37 is the expected volume dependence of 
the spectral weight.}
\label{wt_ratio_m}
\end{figure}
\begin{figure}[htb!]
\includegraphics[height=6.0cm]{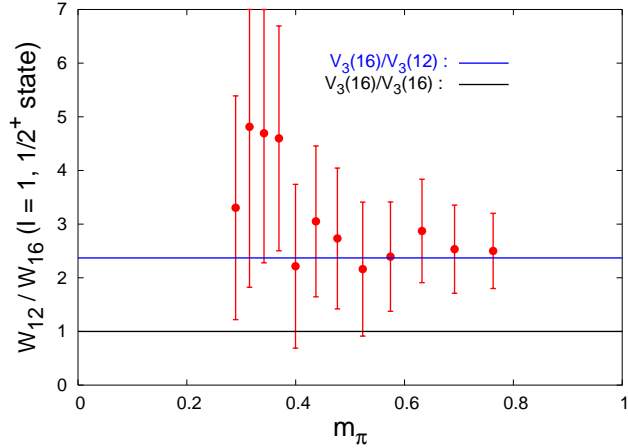}
\caption{Ratio of spectral weights of the two lattices for
$I(J^P)=1\left(1/2^+\right)$ channel. 
The line at 2.37 is the expected volume dependence of the spectral weight.}
\label{wt_ratio_p}
\end{figure}

This volume dependency of spectral weight can be exploited
to check whether the state extracted is a one-particle bound
pentaquark state or a two-particle KN scattering state. This is
especially important to check for the negative parity channel where the
threshold scattering states are close to the presumed pentaquark state.
On the two lattices we used, the ratio of the spectral weights is expected 
to be $W(12)/W(16) \sim 16^3/12^3=2.37$.  Fig.~\ref{wt_ratio_m} shows the ratio
for the negative channel.  Indeed it deviates
significantly from unity, and the value is consistent with $\sim$ 2.37 within
error bars. 
This is strong supporting evidence that the observed state
is a KN scattering state.  The situation is similar in the
positive parity channel, as shown in Fig.~\ref{wt_ratio_p}.  We
conclude from the volume dependence that the ground states we observe
(besides the ghost states) are the KN scattering states, not
the pentaquark states. Similar volume studies have been carried out
earlier~\cite{roper04} to verify that the Roper and $S_{11}$ states
and the ghost $N\eta'$ states do have the expected one-particle and
two-particle volume dependences respectively. Therefore, we caution
that before concluding that a state near the KN scattering threshold
is a pentaquark state, one must study the volume dependence of its
spectral weight.  If the weight has a large volume dependence one
should not mistakenly identify it as the physical pentaquark state. 
The presence of KN scattering states complicates the study of possible
pentaquarks in the 5-quark spectrum.  It is therefore desirable to
have a correlation function which is completely dominated by the
two-particle scattering states and compare it with the point-source
and point-sink correlators which have both the two-particle scattering
states and the possible one-particle pentaquark state. One can also
project out the ground state and excited state explicitly by using
multi-operator cross correlators (similar to Ref.~\cite{csikor}). The
details of such analysis will be presented elsewhere~\cite{penta2}.

\section{Conclusion}
Our results based on the overlap fermion with pion mass as low as
$\sim$ 180 MeV seem to reveal no evidence for a pentaquark state of
the type $uudd\bar{s}$ with the quantum numbers 
$I(J^P)=(0,1)({1\over 2}^\pm)$ near 
a mass of 1540 MeV. Instead, the correlation functions
are dominated by KN scattering states and the ghost $KN\eta'$ states
in the $1/2^+$ channel at low quark mass (pion mass less than 300
MeV).  Our results are consistent with the known features of the KN
scattering phase-shifts analysis~\cite{phase}.  We have checked that
the observed $KN$ states exhibit the expected volume dependence in the
spectral weight for two particles in a box. We advocate the use of
this volume dependence to uncover the character of the states found in
multi-quark calculations on the lattice.

Our conclusion is in contradiction with the other lattice calculations
which have claimed a pentaquark signal of either negative
parity~\cite{csikor,sasaki}, or positive parity~\cite{chiu}, in the
vicinity of 1.54 GeV.


\begin{acknowledgments}
We thank S. Sasaki for useful discussions.
This work is supported in part by U.S. Department of Energy under
grants DE-FG05-84ER40154 and DE-FG02-95ER40907. The computing
resources at NERSC (operated by DOE under DE-AC03-76SF00098) are also
acknowledged.
\end{acknowledgments}

\end{document}